\documentclass[journal]{IEEEtran}

\usepackage[T1]{fontenc}
\usepackage{amsmath,mathtools}
\usepackage[cmintegrals]{newtxmath} 
\usepackage{bm}
\usepackage{graphicx}
\usepackage{algorithm}
\usepackage{algpseudocode}
\usepackage{tabularx,booktabs,multirow}
\usepackage[sort,compress]{cite}
\usepackage{xspace,xcolor}
\usepackage{balance}
\usepackage{subcaption}
\usepackage{tikz}
\usetikzlibrary{arrows.meta,calc,decorations.pathreplacing,shapes.geometric}

\newtheorem{proposition}{Proposition}
\newtheorem{lemma}{Lemma}
\newtheorem{corollary}{Corollary}

\algrenewcommand\algorithmicrequire{\textbf{Input:}}
\algrenewcommand\algorithmicensure{\textbf{Output:}}

\DeclareMathOperator{\tr}{tr}
\DeclareMathOperator{\diag}{diag}

\DeclareMathOperator{\rank}{rank}
\DeclareMathOperator{\sinc}{sinc}
\newcommand{\E}{\mathbb{E}}
\newcommand{\Cbb}{\mathbb{C}}

\newcommand{\bH}{\mathbf{H}}
\newcommand{\bHh}{\widehat{\mathbf{H}}}
\newcommand{\bC}{\mathbf{C}}
\newcommand{\bR}{\mathbf{R}}
\newcommand{\bW}{\mathbf{W}}
\newcommand{\bU}{\mathbf{U}}
\newcommand{\bu}{\mathbf{u}}
\newcommand{\bx}{\mathbf{x}}
\newcommand{\by}{\mathbf{y}}
\newcommand{\bh}{\mathbf{h}}
\newcommand{\ba}{\mathbf{a}}
\newcommand{\bq}{\mathbf{q}}
\newcommand{\bz}{\mathbf{z}}
\newcommand{\bI}{\mathbf{I}}
\newcommand{\bv}{\mathbf{v}}
\newcommand{\bw}{\mathbf{w}}
\newcommand{\bb}{\mathbf{b}}
\newcommand{\bQ}{\mathbf{Q}}
\newcommand{\bB}{\mathbf{B}}
\newcommand{\bG}{\mathbf{G}}
\newcommand{\bL}{\mathbf{L}}
\newcommand{\bT}{\mathbf{T}}
\newcommand{\bP}{\mathbf{P}}
\newcommand{\bV}{\mathbf{V}}
\newcommand{\bX}{\mathbf{X}}
\newcommand{\bY}{\mathbf{Y}}
\newcommand{\bDelta}{\boldsymbol{\Delta}}
\newcommand{\bLambda}{\boldsymbol{\Lambda}}
\newcommand{\bSigma}{\boldsymbol{\Sigma}}
\newcommand{\Dsi}{\mathcal{D}_{\rm SI}}

\newcommand{\Np}{N_{\rm p}}
\newcommand{\Tcpi}{T_{\rm cpi}}
\newif\ifshowrevisions
\showrevisionsfalse   
\ifshowrevisions
  \newcommand{\revcolor}{\color{blue}}
\else
  \newcommand{\revcolor}{}
\fi
\newcommand{\rev}[1]{{\revcolor#1}}       
\newcommand{\revoldcolor}{}                
\newcommand{\revold}[1]{{\revoldcolor#1}}

\begin{document}

\title{Fundamental Limits and Joint Design for Monostatic Joint Communications and Sensing in LEO Satellites}

\author{Chandan~Kumar~Sheemar,~\IEEEmembership{Member,~IEEE,}
        Giovanni~Iacovelli,~\IEEEmembership{Member,~IEEE,}
        Jorge~Querol,~\IEEEmembership{Member,~IEEE,}
        and~Symeon~Chatzinotas,~\IEEEmembership{Fellow,~IEEE}%
\thanks{The authors are with the Interdisciplinary Centre for Security, Reliability and Trust (SnT), University of Luxembourg, L-1855 Luxembourg (e-mail: \{chandan.sheemar, giovanni.iacovelli, jorge.querol, symeon.chatzinotas\}@uni.lu).}%
}

\maketitle

\begin{abstract}
Monostatic joint communication and sensing (JCAS) from a low-Earth-orbit (LEO) satellite requires full-duplex (FD) operation, with self-interference (SI) $80$--$100$~dB stronger than the ground echo. The echo returns milliseconds after transmission, whereas the onboard coupling spans a few symbols. We develop a theory and a design that exploit this separation. Restricting the analog and digital cancellers to the short SI support is without loss of optimality and essential in practice: a least-squares canceller whose support contains the target delay absorbs the coherent echo and reduces the radar SINR by a closed-form factor, about $18$~dB for a vessel at its zero-Doppler crossing. \rev{Modeling the SI that no linear canceller can reconstruct as spatially white} and accounting for ADC quantization, we derive in closed form a ceiling on the radar SINR, the minimum pre-ADC SI power under a canceller tap budget, the budget for full cancellation, and the sensing weight at which the optimal design changes over from communication to sensing. We then maximize \revold{a weighted sum of the communication and sensing SINRs, each normalized by its half-duplex benchmark,} over the beamformers, sensing covariance, combiners, and cancellers for $K$ users and $Q$ targets, using an alternating algorithm with closed-form updates. \revold{Simulations confirm the closed forms. The joint design gains up to $13\%$ in the normalized objective over a sequential design that cancels the SI only at the receiver, and more than $10$~dB of downlink and radar SINR with $4$-bit ADCs and in the 10th-percentile worst-target SINR of four targets, whereas \rev{ignoring the SI overloads the ADCs}.}
\end{abstract}


\section{Introduction}
\label{sec:intro}

\IEEEPARstart{J}{oint} communication and sensing (JCAS), also referred to as integrated sensing and communication, has become a defining capability of sixth-generation networks: a single platform, spectrum, and waveform deliver information and probe the environment at the same time~\cite{liu2022isac,zhang2021overview,liu2020joint,sturm2011waveform}. Non-terrestrial networks are a natural host for this integration~\cite{sheemar2025joint_sur,kodheli2021satellite}. A low-Earth-orbit (LEO) satellite carries a high-gain payload that already covers a wide area from a known position, so the same hardware can provide broadband connectivity and wide-area sensing for maritime surveillance, disaster monitoring, or infrastructure inspection without duplicating radio-frequency \rev{(RF)} chains or spectral resources. In a monostatic implementation, however, the satellite must receive the ground echo while it transmits the downlink over the same time-frequency resources. This is full-duplex (FD) operation~\cite{sheemar2023full,sabharwal2014inband}, and the sensing receiver is exposed to the self-interference (SI) generated by the onboard transmitter, which exceeds the echo of a large vessel by $80$--$100$~dB after passive isolation.

\subsection{State of the Art}
 
FD JCAS has been studied mainly for terrestrial base stations, from FD \rev{orthogonal frequency-division multiplexing (OFDM)} radar measurements~\cite{barneto2019full} to the design of beamformers for FD sensing at millimeter waves~\cite{barneto2022beamformer}, waveforms~\cite{xiao2022waveform}, joint transmit and receive beamformers~\cite{liu2023joint}, joint beamforming and power allocation~\cite{he2023full}, and hybrid architectures~\cite{islam2022isac}. These works model the SI as a short-memory channel whose residual is proportional to the transmit power and treat the target return as contemporaneous with the SI. They rely on the FD cancellation chain of passive suppression~\cite{everett2014passive}, analog and digital cancellation with limited dynamic range~\cite{day2012fullduplex,riihonen2012analog,bharadia2013full}, spatial suppression~\cite{riihonen2011mitigation}, and beamforming-based cancellation at millimeter waves~\cite{roberts2021millimeter,satyanarayana2019hybrid,sheemar2022practical}, and on dual-function waveform and beamforming design without FD constraints~\cite{liu2018mumimo,liu2018toward,tsinos2021joint,hassanien2016signaling,stoica2007probing,li2007mimo}. For satellites, the literature is at the stage of feasibility studies and surveys: FD satellite communications are analyzed in~\cite{lagunas2025full}, JCAS for satellite systems is surveyed in~\cite{sheemar2025joint_sur}, and \rev{massive multiple-input multiple-output (MIMO)} LEO downlinks are designed in~\cite{you2020massive}. On the optimization side, fractional objectives are handled by the quadratic transform~\cite{shen2018fp1,shen2018fp2}, and block-coordinate schemes converge under mild conditions~\cite{razaviyayn2013unified,bertsekas1999nonlinear}.
 
\subsection{Research Gap}
 
Three questions that decide whether monostatic JCAS in space is viable remain open. First, no model separates the delayed echo-generating signal from the contemporaneous SI-generating signal, so the effect of the delay structure on canceller estimation, in particular echo absorption, is unknown. Second, the limits that the SI imposes on spaceborne sensing are not established: how close the radar \rev{signal-to-interference-plus-noise ratio (SINR)} can come to its half-duplex value given the non-reconstructible SI and finite \rev{analog-to-digital converter (ADC)} resolution, which tap budget suffices, and when the payload should switch from communication to sensing. Third, no design jointly optimizes the beamformers, the sensing covariance, the combiners of several targets, and the analog and digital cancellers under the constraints of a satellite payload, and the gain over handling the SI at the receiver only is unknown.
 
\subsection{Contributions}
 
This paper answers these questions and thereby lays the foundations of monostatic JCAS in space. It extends our conference contribution, which introduced the delay-selective architecture for a single target, as follows:
 
\begin{itemize}
\item \emph{Model.} We develop a single-carrier FD LEO JCAS model with $K$ users and $Q$ monostatic targets that makes this separation explicit, describes the coupling by pulse-sampled path responses on a two-sided delay support, and accounts for the motion-induced echo phase, the non-reconstructible SI, and ADC quantization (Section~\ref{sec:model}).
 
\item \emph{Fundamental limits.} We prove that delay-selective cancellers are without loss of optimality (Proposition~\ref{prop:support}), quantify in closed form the echo absorption of a least-squares canceller whose support contains the target delay (Lemma~\ref{lem:nd}, Corollary~\ref{cor:zeta}), and derive closed forms for the radar SINR ceiling \rev{under spatially white non-reconstructible SI, together with a bound for correlated SI} (Proposition~\ref{prop:ceiling}), the minimum pre-ADC SI power and the tap budget for full cancellation (Lemma~\ref{lem:budget}, Corollary~\ref{cor:budget}), and the communication--sensing change-over weight (Proposition~\ref{prop:changeover}) (Section~\ref{sec:analysis}).
 
\item \emph{Joint design.} We maximize the communication and sensing sum-SINRs, each normalized by its half-duplex benchmark and weighted by one scalar per function, over the beamformers, sensing covariance, $Q$ combiners, and both cancellers under power, tap-norm, and pre-ADC constraints by an alternating algorithm with closed-form blocks; both canceller updates and Lemma~\ref{lem:budget} are instances of one generalized Sylvester solution (Lemma~\ref{lem:sylvester}), and \rev{the monotone convergence of the objective} and the complexity are established (Sections~\ref{sec:problem} and~\ref{sec:algorithm}).
 
\item \emph{Evaluation.} With a complete Ka-band link budget, we confirm the closed forms, including the echo loss by a time-domain least-squares Monte Carlo, and compare the joint design with half-duplex references and \rev{with} a sequential design that optimizes the cancellers for the SI-free beamformers\rev{, with an SI-unaware design as an infeasible reference}, under tap budgets, low-resolution ADCs, quantized analog taps, and multiple targets (Section~\ref{sec:simulations}).
\end{itemize}

\subsection{Notation}
Bold letters denote vectors and matrices; $(\cdot)^H$ and $(\cdot)^\dagger$ denote Hermitian transpose and Moore--Penrose inverse; $\tr(\cdot)$, $\|\cdot\|_2$, and $\|\cdot\|_F$ are the trace, Euclidean and Frobenius norms; $\bI_N$ is the $N\times N$ identity; $\delta_{n,m}$ is the Kronecker delta; $\mathcal{CN}(\mathbf{0},\bR)$ is the circularly symmetric complex Gaussian distribution; $\E\{\cdot\}$ and $\Re\{\cdot\}$ are expectation and real part; $\succeq$ is the Loewner order; and $\sinc(x)=\sin(\pi x)/(\pi x)$.

\section{System Model}
\label{sec:model}

\begin{figure*}[t]
\centering
\resizebox{\textwidth}{!}{%
\begin{tikzpicture}[
  font=\footnotesize,
  >={Stealth[length=4pt,width=3pt]},
  blk/.style={draw, rounded corners=1pt, fill=gray!8, minimum height=0.5cm, inner sep=2pt, align=center},
  sum/.style={draw, circle, minimum size=0.32cm, inner sep=0pt},
  ant/.style={draw, isosceles triangle, isosceles triangle apex angle=60, shape border rotate=-90,
              minimum size=0.28cm, inner sep=0pt, fill=white},
  sig/.style={->, thick},
  si/.style={->, thick, red!75!black},
  echo/.style={->, thick, dashed, orange!85!black}
]
\fill[green!6] (0,0.25) -- (0,0.72) .. controls (1.8,1.02) and (3.1,1.08) .. (3.9,1.06) -- (3.9,0.25) -- cycle;
\fill[blue!8]  (3.9,0.25) -- (3.9,1.06) .. controls (4.8,1.04) and (5.7,0.92) .. (6.4,0.74) -- (6.4,0.25) -- cycle;
\draw[gray!60] (0,0.72) .. controls (1.8,1.02) and (3.1,1.08) .. (3.9,1.06) .. controls (4.8,1.04) and (5.7,0.92) .. (6.4,0.74);
\coordinate (tx) at (2.98,4.13);
\coordinate (rx) at (3.46,4.13);
\fill[gray!30, draw=gray!70] (2.85,4.30) rectangle (3.59,4.85);
\foreach \s/\a/\b in {-1/1.55/2.73, 1/3.71/4.89}{
  \fill[blue!18, draw=blue!45] (\a,4.43) rectangle (\b,4.72);
  \foreach \x in {0.25,0.5,...,1.0} \draw[blue!45] ($(\a,4.43)+(\x*1.18,0)$) -- ++(0,0.29);
}
\draw[gray!70] (2.73,4.575) -- (2.85,4.575) (3.59,4.575) -- (3.71,4.575);
\foreach \x in {-0.12,0,0.12} \fill[blue!60!black] ($(tx)+(\x-0.045,0.10)$) rectangle ++(0.09,0.07);
\foreach \x in {-0.12,0,0.12} \fill[green!45!black] ($(rx)+(\x-0.045,0.10)$) rectangle ++(0.09,0.07);
\node[above, align=center] at (3.22,4.85) {LEO satellite, $h=300$~km};
\node[left, blue!60!black, font=\scriptsize] at ($(tx)+(-0.15,0.14)$) {Tx, $N_t$};
\node[right, green!45!black, font=\scriptsize] at ($(rx)+(0.15,0.14)$) {Rx, $N_r$};
\draw[si] ($(tx)+(0.04,0.06)$) .. controls ($(tx)+(0.12,-0.32)$) and ($(rx)+(-0.12,-0.32)$) .. ($(rx)+(-0.04,0.06)$);
\node[red!75!black, font=\scriptsize, align=center] at (3.22,3.62) {SI: $\mathbf{H}_{\rm SI}[d]$, $d\in\mathcal{D}_{\rm SI}$};
\foreach \i/\x/\y/\l in {1/0.65/0.87/{UE\,1}, 2/1.45/0.99/{UE\,2}, 3/2.30/1.06/{UE\,$K$}}{
  \fill[blue!22, opacity=0.55] (tx) -- (\x-0.17,\y+0.22) -- (\x+0.17,\y+0.22) -- cycle;
  \draw[thick] (\x,\y) -- (\x,\y+0.16);
  \draw[thick, fill=white] (\x-0.07,\y+0.24) arc (180:360:0.07 and 0.05) -- cycle;
  \node[below, font=\scriptsize] at (\x,\y+0.02) {\l};
}
\node[blue!60!black, font=\scriptsize] at (1.05,2.35) {$\mathbf{w}_k s_k[n]$};
\coordinate (ship) at (5.05,1.04);
\fill[orange!30, opacity=0.6] (tx) -- ($(ship)+(-0.30,0.26)$) -- ($(ship)+(0.30,0.26)$) -- cycle;
\fill[gray!65] ($(ship)+(-0.36,0)$) -- ($(ship)+(0.38,0)$) -- ($(ship)+(0.28,-0.13)$) -- ($(ship)+(-0.28,-0.13)$) -- cycle;
\fill[gray!45] ($(ship)+(-0.12,0)$) rectangle ++(0.22,0.14);
\node[below, font=\scriptsize] at ($(ship)+(0,-0.14)$) {target $q$};
\draw[echo] ($(ship)+(0.22,0.24)$) .. controls (5.15,2.6) and (4.2,3.7) .. ($(rx)+(0.05,-0.02)$);
\node[orange!85!black, font=\scriptsize, align=left, anchor=west] at (4.95,2.55) {echo of $\mathbf{x}[n-d_q]$,\\ $\tau_q=2R_q/c\approx2.3$~ms};
\node[orange!80!black, font=\scriptsize] at (3.90,1.85) {$\mathbf{R}_x$};
\node at (3.2,-0.05) {(a)};

\begin{scope}[xshift=6.75cm]
  \node (x) at (0.25,4.55) {$\mathbf{x}[n]$};
  \node[blk] (pa) at (1.65,4.55) {RRC, PA};
  \node[ant, fill=blue!15] (tant) at (2.95,4.55) {};
  \node[above, font=\scriptsize] at (tant.north) {Tx};
  \draw[sig] (x) -- (pa);
  \draw[sig] (pa) -- (tant.west);
  \node[ant, fill=green!15] (rant) at (4.75,4.55) {};
  \node[above, font=\scriptsize] at (rant.north) {Rx};
  \draw[si] (tant.east) .. controls (3.55,4.95) and (4.15,4.95) .. (rant.west);
  \node[red!75!black, font=\scriptsize] at (3.85,5.07) {SI};
  \draw[echo] (5.55,4.95) node[above, font=\scriptsize, orange!85!black] {echoes} -- ++(-0.45,-0.25) -- (rant.east);
  \node[sum] (sa) at (4.75,3.55) {$+$};
  \node[blk, fill=green!10] (adc) at (4.75,2.55) {LNA, ADC\\[-1pt]$b$ bits};
  \node[sum] (sd) at (4.75,1.55) {$+$};
  \node[blk, fill=orange!12] (rc) at (4.55,0.55) {range compression~\eqref{eq:rc}\\[-1pt]and combiner: $\mathbf{u}_q^H\mathbf{r}_q$};
  \draw[sig] (rant.south) -- (sa);
  \draw[sig] (sa) -- (adc);
  \draw[sig] (adc) -- (sd);
  \draw[sig] (sd) -- (sd|-rc.north);
  \draw[thick] (0.25,4.32) -- (0.25,0.55);
  \node[blk, fill=red!8] (ca) at (2.35,3.55) {analog $\mathbf{C}_{\rm A}[d]$\\[-1pt]$d\in\mathcal{D}_{\rm SI}$};
  \node[blk, fill=red!8] (cd) at (2.35,1.55) {digital $\mathbf{C}_{\rm D}[d]$\\[-1pt]$d\in\mathcal{D}_{\rm SI}$};
  \draw[sig] (0.25,3.55) -- (ca);
  \draw[sig] (0.25,1.55) -- (cd);
  \draw[sig] (ca) -- node[above, pos=0.75] {$-$} (sa);
  \draw[sig] (cd) -- node[above, pos=0.75] {$-$} (sd);
  \draw[sig] (0.25,0.55) -- node[above, font=\scriptsize] {delay $d_q$} (rc.west);
  \foreach \y in {3.55,1.55,0.55} \fill (0.25,\y) circle (1.1pt);
  \node at (2.9,-0.05) {(b)};
\end{scope}

\begin{scope}[xshift=13.2cm]
  \draw[->] (0,1.25) -- (4.85,1.25);
  \node[font=\scriptsize, anchor=north] at (3.95,0.88) {delay $d$ (symbols)};
  \draw[->] (0.1,1.15) -- (0.1,4.6) node[above, font=\scriptsize, align=center] {tap power (log)};
  \fill[white] (2.62,1.12) rectangle (2.82,1.38);
  \draw (2.56,1.10) -- (2.68,1.40) (2.76,1.10) -- (2.88,1.40);
  \foreach \k/\h in {-5/0.08,-4/0.12,-3/0.20,-2/0.42,-1/0.88,0/2.25,1/1.95,2/0.95,3/0.48,4/0.25,5/0.15,6/0.10,7/0.06}{
    \draw[red!75!black, very thick] ({1.05+0.14*\k},1.25) -- ++(0,\h);
  }
  \node[below, font=\scriptsize] at (1.05,1.22) {$0$};
  \draw[orange!85!black, very thick] (3.95,1.25) -- ++(0,0.35);
  \fill[orange!85!black] (3.95,1.60) circle (1.2pt);
  \node[below, font=\scriptsize] at (3.95,1.22) {$d_q$};
  \node[orange!85!black, font=\scriptsize, align=center] at (3.95,2.15) {echo\\[-1pt]$d_q\approx4.6\times10^4$};
  \node[red!75!black, font=\scriptsize, anchor=west] at (1.42,3.25) {SI taps};
  \node[gray!30!black, font=\scriptsize, align=center] at (3.45,2.95) {$80$--$100$~dB\\[-1pt]weaker};
  \draw[decorate, decoration={brace, amplitude=4pt, mirror}, green!40!black, thick] (0.33,0.88) -- (2.08,0.88)
    node[midway, below=3pt, font=\scriptsize, align=center] {$\mathcal{D}_{\rm SI}$: delay-selective\\[-1pt]support (no loss)};
  \draw[dashed, gray!65!black, thick] (0.33,3.70) -- (0.33,3.90) -- (4.15,3.90) -- (4.15,3.70);
  \node[gray!25!black, font=\scriptsize, align=center, inner sep=1pt] at (2.35,4.22) {support $\ni d_q$: echo absorbed,\\[-1pt]SINR $\times(1-\zeta_q)$};
  \node at (2.5,-0.05) {(c)};
\end{scope}
\end{tikzpicture}
}
\caption{System overview. (a)~Monostatic FD JCAS from a LEO satellite: the downlink beams serve the users and illuminate the target, whose echo returns after $\tau_q\approx2.3$~ms, while the transmit--receive coupling produces SI. (b)~Delay-selective receiver: the analog and digital cancellers reconstruct the SI from $\mathbf{x}[n]$ on the short-delay support $\mathcal{D}_{\rm SI}$, and only the range-compression reference is delayed by $d_q$. (c)~Delay structure: the SI taps occupy $\mathcal{D}_{\rm SI}$ around $d=0$, the echo lies at $d_q\approx4.6\times10^4$; a canceller whose support contains $d_q$ absorbs the echo (Lemma~\ref{lem:nd}).}
\label{fig:scenario}
\end{figure*}
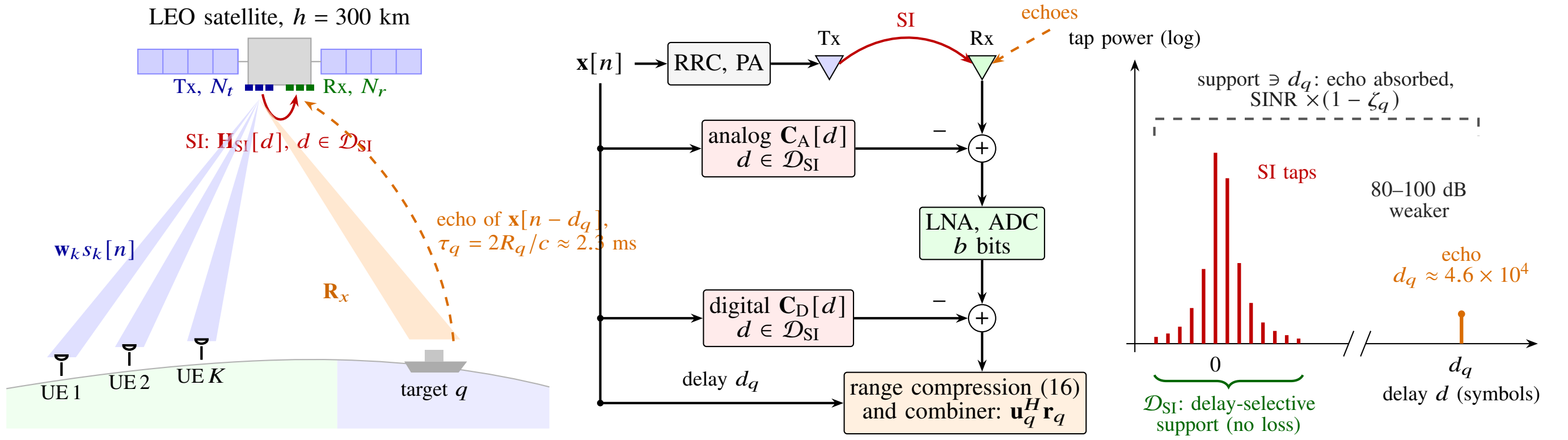

We consider an FD LEO satellite that serves $K$ single-antenna ground users and senses $Q$ ground targets in a monostatic configuration. The payload has $N_t$ transmit and $N_r$ sensing receive antennas (ports of a subarrayed panel), mounted on the same rigid platform. A single-carrier waveform is shared by the communication and sensing functions; it is preferred for its low peak-to-average power ratio at the power-limited satellite amplifier and because the delay separation is explicit at the symbol rate. The support restriction developed below applies equally to the time-domain cancellers of an OFDM transceiver. Fig.~\ref{fig:scenario} illustrates the scenario, the receiver, and the delay structure that the design exploits.

\subsection{Transmit and Communication Models}

Let $\mathcal{K}=\{1,\ldots,K\}$ denote the user set. At symbol time $n$, the satellite transmits
\begin{equation}
\bx[n]=\sum_{k=1}^{K}\bw_ks_k[n]+\bv[n],
\label{eq:tx}
\end{equation}
where $\bw_k\in\Cbb^{N_t}$ is the beamformer of user $k$, $s_k[n]$ is its information symbol, and $\bv[n]\in\Cbb^{N_t}$ is an optional dedicated sensing signal. The symbols are temporally white and mutually independent, $\E\{s_k[n]s_j^*[m]\}=\delta_{k,j}\delta_{n,m}$, and the sensing signal satisfies $\E\{\bv[n]\bv^H[m]\}=\bR_{\rm s}\delta_{n,m}$, $\bR_{\rm s}\succeq\mathbf{0}$, independently of the symbols. The transmit covariance is
\begin{equation}
\bR_x=\E\{\bx[n]\bx^H[n]\}=\sum_{k=1}^{K}\bw_k\bw_k^H+\bR_{\rm s}.
\label{eq:Rx}
\end{equation}
The sequence $\bx[n]$ is pulse-shaped with a fixed root-raised-cosine (RRC) pulse $p(t)$ of symbol period $T_s$, which is known and not optimized. After timing and Doppler compensation, user $k$ receives
\begin{equation}
y_k[n]=\bh_k^H\bw_ks_k[n]+\sum_{j\neq k}\bh_k^H\bw_js_j[n]+\bh_k^H\bv[n]+n_k[n],
\label{eq:user}
\end{equation}
where $\bh_k\in\Cbb^{N_t}$ is the effective satellite-to-user channel and $n_k[n]\sim\mathcal{CN}(0,\sigma_k^2)$. The communication SINR is
\begin{equation}
\gamma_k=\frac{|\bh_k^H\bw_k|^2}{\sum_{j\neq k}|\bh_k^H\bw_j|^2+\bh_k^H\bR_{\rm s}\bh_k+\sigma_k^2}.
\label{eq:sinr_k}
\end{equation}

\subsection{Delayed Target Echoes and SI Channel}
\label{sec:echo}

Target $q\in\mathcal{Q}=\{1,\ldots,Q\}$ has round-trip delay $\tau_q$, integer symbol delay $d_q=\lfloor\tau_q/T_s\rceil$, and fractional delay $\delta_q=\tau_q-d_qT_s$, $|\delta_q|\le T_s/2$, whose effect is a known amplitude factor absorbed into $\beta_q$ (Section~\ref{sec:rc}). The delay $\tau_q$ is predicted from the satellite ephemeris and the target position provided by a preceding detection or tracking stage, or it is a hypothesized range bin in search mode; the design below requires only $d_q\notin\Dsi$ and is otherwise independent of $d_q$, so that errors in $\tau_q$ do not affect it. Within a coherent processing interval (CPI) of duration $\Tcpi$, $d_q$ is held fixed, which requires the range walk to stay below half a symbol, $|\dot r_q|<cT_s/(4\Tcpi)$; this holds near the zero-Doppler instant and, otherwise, after ephemeris-based range-migration compensation. The fast phase variation is retained explicitly.

The target channel is modeled as
\begin{equation}
\bH_q=\beta_q\ba_r(\theta_{q})\ba_t^H(\theta_{q})\in\Cbb^{N_r\times N_t},\quad
|\beta_q|^2=\frac{\psi_p^2(\delta_q)\,G_e^2\lambda_c^2}{(4\pi)^3R_q^4L_a^2},
\label{eq:Hq}
\end{equation}
which collects the two-way propagation loss over the slant range $R_q$, the atmospheric loss $L_a$, the port gain $G_e$, and the straddle factor $\psi_p^2(\delta_q)\le1$ defined below, while $\ba_t(\theta_q)$ and $\ba_r(\theta_q)$ are the transmit and receive steering vectors toward the target. The reflection coefficient $\rho_q$ satisfies $\E\{\rho_q\}=0$ and $\E\{|\rho_q|^2\}=\sigma_q^2$, where $\sigma_q^2$ is the radar cross section (RCS). Satellite motion produces the deterministic carrier phase
\begin{equation}
\phi_q[n]=\exp\big(-j2\pi f_c\tau_q(nT_s)\big),
\label{eq:phase}
\end{equation}
where $f_c$ is the carrier frequency, $\lambda_c=c/f_c$ is the wavelength, and $\tau_q(t)$ follows from the ephemeris. The echo of target $q$ is
\begin{equation}
\by_q[n]=\rho_q\phi_q[n]\bH_q\bx[n-d_q].
\label{eq:echo}
\end{equation}
Hence the echo observed at time $n$ is generated by the transmission emitted $d_q$ symbols earlier, whereas the SI is generated by the symbols transmitted around time $n$ through the short-memory MIMO channel
\begin{equation}
\by_{\rm SI}[n]=\sum_{d\in\Dsi}\bH_{\rm SI}[d]\bx[n-d],\qquad
\Dsi=\{-D_{\rm pre},\ldots,D_{\rm post}\},
\label{eq:si}
\end{equation}
with $D_{\rm SI}=|\Dsi|\ll d_q$. The coupling consists of $L_{\rm SI}$ paths with $N_r\times N_t$ gain matrices $\bG_\ell$ and delays $\tau_\ell$, and the taps are their discrete-time equivalents at the receiver sampling instants of Section~\ref{sec:rc},
\begin{equation}
\bH_{\rm SI}[d]=\sum_{\ell=0}^{L_{\rm SI}-1}\psi_p(dT_s-\tau_\ell)\,\bG_\ell,
\label{eq:si_taps}
\end{equation}
where $\psi_p$ is the pulse autocorrelation defined in Section~\ref{sec:rc}, applied entrywise when $\tau_\ell$ depends on the antenna pair. Since $\tau_\ell\neq0$, the pulse produces precursor taps ($d<0$) as well as tails, e.g., \rev{$22$} and $14$~dB below the main tap for paths at $0.1T_s$ and $0.3T_s$ with roll-off \rev{$0.35$, the value used in Section~\ref{sec:simulations}}, far above any practical cancellation floor. They are realizable because $\bx[n-d]$, $d<0$, has already entered the transmit pulse-shaping filter, and $D_{\rm pre}$ and $D_{\rm post}$ are chosen so that the omitted tail energy lies below the non-reconstructible SI of Section~\ref{sec:cancel}. Since $d_q\ge\lceil2h/(cT_s)\rceil$ for a satellite at altitude $h$, i.e., $d_q\ge4\times10^4$ for $h=300$~km and $1/T_s=20$~MBd, $d_q\notin\Dsi$ holds with a margin of four orders of magnitude. The dominant path, $\ell=0$, is the near-field line-of-sight coupling with the spherical-wave response
\begin{equation}
\big[\bG_0\big]_{r,t}=\sqrt{\kappa_{\rm SI}}\,\frac{\lambda_c}{4\pi d_{r,t}}\exp\Big(-j\frac{2\pi}{\lambda_c}d_{r,t}\Big)
\label{eq:nearfield}
\end{equation}
and delay $\tau_{0,r,t}=d_{r,t}/c$, where $d_{r,t}$ is the distance between transmit antenna $t$ and receive antenna $r$, and $0<\kappa_{\rm SI}\le1$ is the passive-isolation attenuation. The remaining paths model short-delay reflections from the satellite structure.

\subsection{Delay-Selective Cancellation With Impairments}
\label{sec:cancel}

The analog canceller subtracts $\mathbf{c}_{\rm A}[n]=\sum_{d\in\Dsi}\bC_{\rm A}[d]\bx[n-d]$ before the low-noise amplifiers and ADCs, and the digital canceller subtracts $\mathbf{c}_{\rm D}[n]=\sum_{d\in\Dsi}\bC_{\rm D}[d]\bx[n-d]$ after sampling, with $\bC_{\rm A}[d],\bC_{\rm D}[d]\in\Cbb^{N_r\times N_t}$. Both are computed from an estimate $\bHh_{\rm SI}[d]$ of the coupling taps, obtained by least squares on the SI support from previous CPIs or from a calibration sequence (Lemma~\ref{lem:nd}(a) in Section~\ref{sec:analysis})\revold{; samples from previous CPIs identify the taps only on the range of the transmit covariance used in them, whereas a full-rank calibration sequence identifies them in every transmit direction, as the tie-breaking rule of Section~\ref{sec:tiebreak} requires}. For $\Np\gg N_tD_{\rm SI}$ its estimation error lies well below the non-reconstructible SI introduced next, so we set $\bHh_{\rm SI}[d]=\bH_{\rm SI}[d]$ and keep the hat for the quantities available to the cancellers. What limits the cancellation is the part of the SI that no linear canceller over $\Dsi$ can reconstruct: transmitter noise, power-amplifier distortion, phase noise, and coupling drift between estimation and use~\cite{day2012fullduplex}. We model it as a zero-mean, time-varying perturbation of the taps, so that the received SI is
\begin{equation}
\by_{\rm SI}[n]=\sum_{d\in\Dsi}\big(\bH_{\rm SI}[d]+\bDelta_n[d]\big)\bx[n-d],
\label{eq:estimate}
\end{equation}
where $\bDelta_n[d]$ is independent of $\bx$ and of $\bHh_{\rm SI}[d]$ and has i.i.d.\ $\mathcal{CN}(0,\sigma_{\Delta,d}^2)$ entries with $\sigma_{\Delta,d}^2=\epsilon^2\|\bH_{\rm SI}[d]\|_F^2/(N_rN_t)$. Its temporal correlation does not affect the second-order quantities below, and, being uncorrelated with $\bx[n-d]$, it is not removed by any least-squares fit. The parameter $\epsilon^2$ is the normalized power of the non-reconstructible SI, which we call the SI-knowledge floor, and $\sigma_\Delta^2=\sum_d\sigma_{\Delta,d}^2$. The residual channels after the analog and after both stages are
\begin{align}
\bHh_{\rm A}[d]&=\bHh_{\rm SI}[d]-\bC_{\rm A}[d],\label{eq:HA}\\
\bHh_{\rm res}[d]&=\bHh_{\rm SI}[d]-\bC_{\rm A}[d]-\bC_{\rm D}[d],\label{eq:Hres}
\end{align}
and the true residuals are $\bHh_{\rm A}[d]+\bDelta_n[d]$ and $\bHh_{\rm res}[d]+\bDelta_n[d]$. Averaging over $\bDelta_n[d]$ under the whiteness assumptions, the covariance of the SI entering the ADCs is
\begin{equation}
\bR_{\rm A}^{\rm SI}=\sum_{d\in\Dsi}\bHh_{\rm A}[d]\bR_x\bHh_{\rm A}^H[d]+\sigma_\Delta^2\tr(\bR_x)\bI_{N_r}.
\label{eq:RA}
\end{equation}
For $b$-bit I/Q ADCs with automatic gain control, clipped at the peak of the complex envelope, the signal-to-quantization-noise ratio is $\mathrm{SQNR}=6.02b+1.76-\mathrm{PAPR}$~[dB], with \rev{PAPR the peak-to-average power ratio} of the complex envelope (equivalently, $6.02b+4.77$~dB minus the PAPR of each real component~\cite{riihonen2012analog}). Using the branch-averaged input power, the quantization noise in each receive branch has power $\varrho\tr(\bR_{\rm A}^{\rm SI})$ with $\varrho=10^{-\mathrm{SQNR}/10}/N_r$~\cite{day2012fullduplex}. The post-cancellation signal is therefore
\begin{equation}
\by_{\rm D}[n]=\sum_{q}\by_q[n]+\sum_{d\in\Dsi}(\bHh_{\rm res}[d]+\bDelta_n[d])\bx[n-d]+\bz_r[n]+\bz_{\rm Q}[n],
\label{eq:yD}
\end{equation}
where $\bz_r[n]\sim\mathcal{CN}(\mathbf{0},\sigma_r^2\bI_{N_r})$ is thermal noise and $\bz_{\rm Q}[n]$ is the quantization noise. Both cancellers are \emph{delay selective}: their nonzero taps are restricted to $\Dsi$, so they cannot generate a replica at any target delay.

\subsection{Range Compression and Delay-Resolved Radar SINR}
\label{sec:rc}

Let $\psi_p(\tau)$ denote the normalized autocorrelation of the RRC pulse, i.e., the raised-cosine response after receive matched filtering, with $\psi_p(kT_s)=\delta_{k,0}$. The receiver samples the matched-filter output once per symbol at $t_n=nT_s$, common to all targets. The return of target $q$ then contributes $\rho_q\phi_q[n]\bH_q\sum_i\psi_p((n-d_q-i)T_s-\delta_q)\bx[i]$, whose dominant term is~\eqref{eq:echo} scaled by $\psi_p(\delta_q)$, absorbed into $\beta_q$ in~\eqref{eq:Hq}, with a straddle loss of at most about $4$~dB and about $1.2$~dB on average over $\delta_q$; the neighboring terms are uncorrelated with the range-compression reference and are negligible. Likewise, an SI path with delay $\tau_\ell$ contributes $\sum_i\psi_p((n-i)T_s-\tau_\ell)\bG_\ell\bx[i]$, including precursors from $i>n$, which gives the taps~\eqref{eq:si_taps}, the same for every target since the sampling grid is common. Because the satellite transmits continuously, the SI in the samples of the delay bin of target $q$ is generated by the symbols $\bx[n-d]$, $d\in\Dsi$, emitted around each sampling instant, and not by $\bx[n-d_q]$.

The target is detected by range compression over the $\Np=\Tcpi/T_s$ samples of a CPI. With the known illumination sequence $g_q[n]=\ba_t^H(\theta_q)\bx[n]$ and $P_q=\ba_t^H(\theta_q)\bR_x\ba_t(\theta_q)$, the range- and Doppler-matched output at the delay bin of target $q$ is
\begin{equation}
\mathbf{r}_q=\frac{1}{\sqrt{\Np P_q}}\sum_{n=1}^{\Np}\phi_q^*[n]\,g_q^*[n-d_q]\,\by_{\rm D}[n].
\label{eq:rc}
\end{equation}
Since $\bx[n-d_q]$ is independent of $\bx[n-d]$ for $d\in\Dsi$ and of $\bx[n-d_{q'}]$ for $q'\neq q$ \revold{(targets in different range bins, $d_{q'}\neq d_q$)}, the echo of target $q$ adds coherently in~\eqref{eq:rc}, whereas the residual SI, the SI-knowledge error, the thermal and quantization noise, and the echoes of the other targets are uncorrelated with the reference and are integrated incoherently in the same way. For $\Np\gg1$,
\begin{equation}
\mathbf{r}_q\approx\sqrt{\Np P_q}\,\rho_q\beta_q\ba_r(\theta_q)+\mathbf{i}_q,
\label{eq:rq}
\end{equation}
with target covariance $\Np\bR_q$, $\bR_q=\sigma_q^2\bH_q\bR_x\bH_q^H$, and interference-plus-noise covariance
\begin{align}
\bR_{\rm i}&=\bR_{\rm SI}^{\rm res}+\big(\sigma_r^2+\varrho\tr(\bR_{\rm A}^{\rm SI})\big)\bI_{N_r},\label{eq:Ri}\\
\bR_{\rm SI}^{\rm res}&=\sum_{d\in\Dsi}\bHh_{\rm res}[d]\bR_x\bHh_{\rm res}^H[d]+\sigma_\Delta^2\tr(\bR_x)\bI_{N_r},\label{eq:Rres}
\end{align}
where the echoes of the other targets, whose per-sample power is $50$~dB below the noise in the scenarios of interest, have been neglected in $\bR_{\rm i}$. The processing gain $\Np$ thus multiplies the target term only, while residual SI and noise are treated identically. With a receive combiner $\bu_q\in\Cbb^{N_r}$ per target, the delay-resolved radar SINR is
\begin{equation}
\gamma_q=\frac{\Np\,\bu_q^H\bR_q\bu_q}{\bu_q^H\bR_{\rm i}\bu_q},\qquad q\in\mathcal{Q}.
\label{eq:sinr_q}
\end{equation}
Note that $\bR_{\rm i}$ is common to all targets: the cancellers and the transmit covariance shape one interference field that every combiner sees, and only the reference in~\eqref{eq:rc} and the combiner $\bu_q$ are target specific.

\section{Fundamental Properties}
\label{sec:analysis}

This section establishes the results that govern the design: the support restriction is without loss of optimality; a canceller estimated over a support that contains the target delay absorbs the echo by a computable factor; the radar SINR has a closed-form ceiling set by the non-reconstructible SI\rev{, here modeled as spatially white}; the pre-ADC SI power achievable under a tap budget has a closed form; and the optimal design changes over from communication to sensing at a closed-form weight. Proofs are given in the Appendix.

\subsection{Delay Selectivity Is Without Loss of Optimality}

Let $\mathcal{J}(\bW,\{\bu_q\},\{\bC_{\rm A}[d]\},\{\bC_{\rm D}[d]\})=\sum_k\alpha_k\gamma_k+\sum_q\alpha_q\gamma_q$ denote the weighted sum-SINR with nonnegative weights, evaluated with~\eqref{eq:sinr_k} and~\eqref{eq:sinr_q}, where the sums in~\eqref{eq:RA} and~\eqref{eq:Rres} now run over every delay at which a canceller tap is nonzero.

\begin{proposition}[Support invariance]
\label{prop:support}
Let $\{\bC_{\rm A}[d],\bC_{\rm D}[d]\}_{\revold{d\in\mathbb{Z}}}$ be any cancellers with finite support, and let $\{\bC'_{\rm A}[d],\bC'_{\rm D}[d]\}$ be obtained by setting to zero every tap with $d\notin\Dsi$. Then, for every $\bW$ and $\{\bu_q\}$, the primed cancellers satisfy the tap-norm and pre-ADC constraints whenever the unprimed ones do, and $\mathcal{J}$ does not decrease. Consequently, the delay-support restriction $\bC_{\rm A}[d]=\bC_{\rm D}[d]=\mathbf{0}$, $d\notin\Dsi$, can be imposed without loss of optimality in any design problem whose objective is nondecreasing in each $\gamma_q$ and $\gamma_k$.
\end{proposition}

The proposition states that delay selectivity adds no degree of freedom: a tap at $d\notin\Dsi$ can only add a positive semidefinite term to the interference, coherently integrated if $d=d_q$, consume budget, and increase the pre-ADC power. Its role is therefore not to improve the optimum but to make the optimum reachable when the cancellers are estimated from data, which is the subject of the next result.

\subsection{Echo Absorption by a Non-Delay-Selective Canceller}

The estimate $\bHh_{\rm SI}[d]$ from which the cancellers are computed is itself obtained by regressing received samples on delayed transmit symbols, and the echo is present in those samples. The next result shows that the support of that regression decides whether the echo survives. Consider a least-squares (LS) canceller with support $\mathcal{S}\subset\mathbb{Z}$, estimated from the $\Np$ post-analog samples $\bY=[\by[1],\ldots,\by[\Np]]$ and the stacked regressors $\bX=[\bX_{d}]_{d\in\mathcal{S}}$, $\bX_d=[\bx[1-d],\ldots,\bx[\Np-d]]$, as $[\bC[d]]_{d\in\mathcal{S}}=\bY\bX^\dagger$, and applied to the same block.

\begin{lemma}[Echo absorption]
\label{lem:nd}
Let $\bx[n]$ be temporally white with covariance $\bR_x$, and let $\bar\phi_q=\Np^{-1}\sum_{n=1}^{\Np}\phi_q[n]$ and $\zeta_q=|\bar\phi_q|^2\in[0,1]$ denote the coherence of the echo phase over the CPI. As $\Np\to\infty$:
\begin{enumerate}
\item[(a)] if $\mathcal{S}\subseteq\Dsi$, then $\bC[d]\to\bHh_{\rm A}[d]\bR_x\bR_x^\dagger$ for $d\in\mathcal{S}$, which removes the reconstructible post-analog SI generated by $\bx[n]$ on $\mathcal{S}$, leaves $\bDelta_n[d]$, and leaves the range-compressed echo unchanged;
\item[(b)] if $d_q\in\mathcal{S}$, then $\bC[d_q]\to\rho_q\bar\phi_q\bH_q\bR_x\bR_x^\dagger$, and the range-compressed output equals that of the samples without the fitted taps, compressed with the reference $\phi_q^*[n]-\bar\phi_q^*$ instead of $\phi_q^*[n]$. The echo power is scaled by $(1-\zeta_q)^2$ and the power of the residual SI and noise by $1-\zeta_q$, so that
\begin{equation}
\gamma_q^{\rm ND}=(1-\zeta_q)\,\gamma_q\,\big(1+\mathcal{O}(|\mathcal{S}|N_t/\Np)\big),
\label{eq:nd_loss}
\end{equation}
where $\gamma_q$ is the SINR obtained with the same $(\bR_x,\bu_q)$ and the delay-selective canceller of (a) on $\mathcal{S}\cap\Dsi$.
\end{enumerate}
\end{lemma}

\begin{corollary}[Coherence factor in closed form]
\label{cor:zeta}
For a Doppler offset \revold{$f_{\rm D}$ of the echo in the samples from which the canceller is estimated}, $\phi_q[n]=e^{j2\pi f_{\rm D}nT_s}$, and $\Np\gg1$,
\begin{equation}
\zeta_q=\sinc^2(f_{\rm D}\Tcpi).
\label{eq:zeta_lin}
\end{equation}
Around the zero-Doppler instant, $\phi_q(t)=\revold{e^{-j\pi f_{\rm R}t^2}}$ for $t\in[-\Tcpi/2,\Tcpi/2]$ with Doppler rate $f_{\rm R}=2v_{\rm eff}^2/(\lambda_cR_q)$, and
\begin{equation}
\zeta_q=\frac{C^2(X)+S^2(X)}{X^2},\qquad X=\sqrt{\frac{f_{\rm R}\Tcpi^2}{2}},
\label{eq:zeta_chirp}
\end{equation}
where $C(\cdot)$ and $S(\cdot)$ are the Fresnel cosine and sine integrals, and $v_{\rm eff}$ is the effective relative velocity.
\end{corollary}

Lemma~\ref{lem:nd} makes precise what the separation buys. The SI channel has no energy at $d_q$, so the tap that a non-delay-selective canceller places there is made entirely of echo, whose coherent part $\bar\phi_q$ is fitted and removed. Because the fit is orthogonal to the regressor $\bx[n-d_q]$, which is also the range-compression reference, the residual SI and noise in the target bin are reduced in the same proportion as the remaining echo amplitude, so the SINR loss is $1-\zeta_q$. Near the zero-Doppler crossing $\zeta_q$ is close to one: for $f_{\rm R}=2.2\times10^4$~Hz/s and $\Tcpi=5$~ms, $\zeta_q=0.983$ and the loss is $17.8$~dB. For $f_{\rm D}\Tcpi\ge1$\revold{, i.e., $|t|\ge1/(f_{\rm R}\Tcpi)\approx9$~ms from the zero-Doppler instant since $f_{\rm D}\approx f_{\rm R}t$,} the echo decorrelates within the CPI and the loss vanishes, but the Doppler compensation in~\eqref{eq:rc} and the range-walk correction are then mandatory. The delay-selective canceller of part (a) is immune in all cases. \rev{Lemma~\ref{lem:nd} concerns a canceller estimated on the block to which it is applied, as in on-line estimation from the operating signal, which is also the setting of Fig.~\ref{fig:ls}. If the taps are estimated on other data, the tap at $d_q$ reflects the echo contained in those data: it absorbs the current echo to the extent that the two are coherent, e.g., for a target observed again with the same reflection coefficient and phase history, and otherwise subtracts a replica uncorrelated with the current echo, which only adds interference (Proposition~\ref{prop:support}). An echo-free calibration sequence, or the delay-selective support, avoids both effects.}

\subsection{A Closed-Form Ceiling on the Radar SINR}\label{sec:ceil}

Define the interference-to-noise ratio per receive branch under isotropic transmission at full power,
\begin{equation}
\mathrm{INR}=\frac{P_{\max}\sum_{d\in\Dsi}\|\bH_{\rm SI}[d]\|_F^2}{N_rN_t\sigma_r^2},
\label{eq:inr}
\end{equation}
so that $\sigma_\Delta^2P_{\max}/\sigma_r^2=\epsilon^2\,\mathrm{INR}$, and the half-duplex (HD) sensing \rev{signal-to-noise ratio (SNR)} of target $q$,
\begin{equation}
\bar\gamma_q=\frac{\Np\sigma_q^2|\beta_q|^2N_rN_tP_{\max}}{\sigma_r^2},
\label{eq:hd_radar}
\end{equation}
obtained by beamforming $P_{\max}$ toward the target without SI.

\begin{proposition}[Radar SINR ceiling]
\label{prop:ceiling}
\rev{Under the model~\eqref{eq:estimate} of the non-reconstructible SI,} for every transmit covariance with $\tr(\bR_x)\le P_{\max}$, every unit-norm combiner, and all \rev{linear} cancellers,
\begin{equation}
\gamma_q\le\gamma_q^{\star}\triangleq\frac{\bar\gamma_q}{1+\epsilon^2\,\mathrm{INR}\,(1+\varrho N_r)}.
\label{eq:ceiling}
\end{equation}
The bound is attained by $\bR_x=P_{\max}\ba_t\ba_t^H/N_t$, $\bu_q=\ba_r/\sqrt{N_r}$, $\bC_{\rm D}[d]=\mathbf{0}$, and $\bC_{\rm A}[d]=\bHh_{\rm SI}[d]\ba_t\ba_t^H/N_t$, provided that $\sum_d\|\bHh_{\rm SI}[d]\ba_t\|_2^2/N_t\le P_{\rm A}^{\max}$ and $N_r\sigma_\Delta^2P_{\max}\le P_{\rm ADC}^{\max}$.
\end{proposition}

\rev{Under this model,} the ceiling depends on the SI only through the product $\epsilon^2\,\mathrm{INR}$: the reconstructible part of the SI can always be removed from the target bin, so the sensing performance of an FD LEO payload is limited by the power of the SI that no linear canceller can reconstruct, $\epsilon^2\,\mathrm{INR}$, and not by the reconstructible coupling, however strong it is. For $\mathrm{INR}=47$~dB and $\epsilon^2=-50$~dB, the gap to the HD benchmark is $10\log_{10}(1+10^{-0.3})=1.8$~dB; with $\epsilon^2=-30$~dB it is $17$~dB. Since $\varrho N_r=10^{-\mathrm{SQNR}/10}\ll1$ for any practical ADC, quantization does not affect the ceiling; it affects feasibility through the pre-ADC constraint, which is the subject of the next result.

\rev{The ceiling is thus a property of the impairment model, not of the architecture alone. In~\eqref{eq:estimate}, the non-reconstructible SI is spatially white at the receiver and proportional to $\tr(\bR_x)$, so no combiner can avoid it. If its contribution to the target bin has instead a general covariance $\bSigma_\Delta(\bR_x)\succeq\mathbf{0}$, e.g., for transmitter noise radiated through the coupling channel, dropping the reconstructible residual and the quantization noise and maximizing over $\bu_q$ as in the proof gives
\[
\gamma_q\le\Np\sigma_q^2|\beta_q|^2\max_{\tr(\bR_x)\le P_{\max}}\big(\ba_t^H\bR_x\ba_t\big)\,\ba_r^H\big(\sigma_r^2\bI_{N_r}+\bSigma_\Delta(\bR_x)\big)^{-1}\ba_r,
\]
which reduces to~\eqref{eq:ceiling}, up to the quantization factor, for $\bSigma_\Delta(\bR_x)=\sigma_\Delta^2\tr(\bR_x)\bI_{N_r}$. A correlated floor can raise the ceiling, when the combiner can steer away from it, or lower it, when it concentrates along $\ba_r(\theta_q)$; Section~\ref{sec:simulations} uses the spatially white model.}

\subsection{Pre-ADC SI Power Under a Tap Budget}

Let $\bR_x=\sum_{j=1}^{N_t}p_j\mathbf{f}_j\mathbf{f}_j^H$ be the eigendecomposition of the transmit covariance, with $p_j\ge0$ the power in eigendirection $\mathbf{f}_j$, and let $h_j=\sum_{d\in\Dsi}\|\bHh_{\rm SI}[d]\mathbf{f}_j\|_2^2$ be the SI-channel energy in that direction.

\begin{lemma}[Minimum pre-ADC SI power]
\label{lem:budget}
The minimum of $\Pi_{\rm A}=\sum_{d\in\Dsi}\tr\big(\bHh_{\rm A}[d]\bR_x\bHh_{\rm A}^H[d]\big)$ over analog cancellers with $\sum_d\|\bC_{\rm A}[d]\|_F^2\le P_{\rm A}^{\max}$ is
\begin{equation}
\Pi_{\rm A}^{\min}(P_{\rm A}^{\max})=\sum_{j=1}^{N_t}\frac{\vartheta^2p_jh_j}{(p_j+\vartheta)^2},
\label{eq:pimin}
\end{equation}
attained by $\bC_{\rm A}[d]=\bHh_{\rm SI}[d]\bR_x(\bR_x+\vartheta\bI_{N_t})^{-1}$ \revold{(read as $\bHh_{\rm SI}[d]\bR_x\bR_x^\dagger$ if $\vartheta=0$ and $\bR_x$ is singular, in which case the minimizer is not unique)}, where $\vartheta=0$ if $\sum_{j:p_j>0}h_j\le P_{\rm A}^{\max}$ and otherwise $\vartheta>0$ is the unique root of $\sum_jp_j^2h_j/(p_j+\vartheta)^2=P_{\rm A}^{\max}$.
\end{lemma}

\begin{corollary}[Budget for full cancellation]
\label{cor:budget}
The known SI can be removed completely before the ADCs if and only if
\begin{equation}
P_{\rm A}^{\max}\ge P_{\rm A}^\star(\bR_x)\triangleq\sum_{j:p_j>0}h_j.
\label{eq:budget_star}
\end{equation}
If $\rank(\bR_x)=r$ and the eigenvectors are isotropically distributed and independent of $\bHh_{\rm SI}$, then $\E\{P_{\rm A}^\star(\bR_x)\}=(r/N_t)\sum_d\|\bHh_{\rm SI}[d]\|_F^2$, i.e., a relative tap budget $\xi_{\rm A}=P_{\rm A}^{\max}/\sum_d\|\bHh_{\rm SI}[d]\|_F^2$ of about $r/N_t$ suffices on average; for the structured near-field coupling of~\eqref{eq:nearfield} the realization-dependent factor lies between \revold{$0.37$ and $1.61$} in the scenarios of Section~\ref{sec:simulations}.
\end{corollary}

Corollary~\ref{cor:budget} is the practical companion of Proposition~\ref{prop:ceiling}. At the sensing-optimal point the transmit covariance has rank one, so a budget of about $1/N_t$ of the SI-channel energy suffices to reach the ceiling; the exact condition of Proposition~\ref{prop:ceiling}, $\sum_d\|\bHh_{\rm SI}[d]\ba_t\|_2^2/N_t\le P_{\rm A}^{\max}$, holds with $\xi_{\rm A}=0.1$ in \revold{$100\%$} of the realizations of Section~\ref{sec:simulations} for $N_t=16$ but only in \revold{$40\%$} for $N_t=8$, and a rank-$Q$ sensing covariance for $Q$ targets needs about $Q/N_t$. When the budget is insufficient, $\Pi_{\rm A}^{\min}>0$ and the design must either accept residual SI in the target bin or steer the transmit covariance away from the strongly coupled eigendirections, which costs illumination power. Lemma~\ref{lem:budget} also settles the cost of FD operation on the downlink: \revold{if $\Pi_{\rm A}^{\min}(P_{\rm A}^{\max})+N_r\sigma_\Delta^2P_{\max}\le P_{\rm ADC}^{\max}$ at the HD-optimal communication covariance}, the HD beamformers are feasible for the FD problem and FD sensing costs nothing on the communication side at low sensing weights, as confirmed in Section~\ref{sec:simulations}.

\subsection{Communication--Sensing Change-Over}
\label{sec:changeover}

The weighted sum-SINR is linear in the SINRs, and once the reconstructible SI is cancelled the SINRs are linear in the power given to each function. The optimal design therefore switches abruptly between communication and sensing, at a weight that has a closed form. Let $Q=1$ and \revold{weigh the two functions by $\alpha_{\rm c},\alpha_{\rm r}\ge0$, $\alpha_{\rm c}+\alpha_{\rm r}=1$, each normalized by its HD benchmark, i.e., consider the objective $\alpha_{\rm c}\sum_k\gamma_k/\bar\Gamma_{\rm c}+\alpha_{\rm r}\gamma_q/\bar\gamma_q$ adopted in Section~\ref{sec:problem}}, where $\bar\Gamma_{\rm c}$ is the HD communication sum-SINR, i.e., the maximum of $\sum_k\gamma_k$ without sensing and without SI, and let
\begin{equation}
\bar\rho=\frac{\gamma_q^\star}{\bar\gamma_q}=\frac{1}{1+\epsilon^2\,\mathrm{INR}\,(1+\varrho N_r)}
\label{eq:rho_bar}
\end{equation}
be the normalized ceiling of Proposition~\ref{prop:ceiling}.

\begin{proposition}[Change-over weight]
\label{prop:changeover}
Let the user channels be mutually orthogonal and orthogonal to the target, $\bh_k^H\bh_j=0$ for $k\neq j$ and $\bh_k^H\ba_t(\theta_q)=0$. Then every feasible point has normalized objective \revold{$\alpha_{\rm c}\sum_k\gamma_k/\bar\Gamma_{\rm c}+\alpha_{\rm r}\gamma_q/\bar\gamma_q\le\max\{\alpha_{\rm c},\alpha_{\rm r}\bar\rho\}$}. If, in addition, the HD-optimal communication beamformers satisfy the pre-ADC constraint and the conditions of Proposition~\ref{prop:ceiling} hold, the bound is attained, by serving only the users for \revold{$\alpha_{\rm r}<\alpha_{\rm r}^\star$} and only the target for \revold{$\alpha_{\rm r}>\alpha_{\rm r}^\star$}, where
\begin{equation}
{\revoldcolor\alpha_{\rm r}^\star}=\frac{1}{1+\bar\rho}=\frac{1+\epsilon^2\,\mathrm{INR}\,(1+\varrho N_r)}{2+\epsilon^2\,\mathrm{INR}\,(1+\varrho N_r)}.
\label{eq:alpha_star}
\end{equation}
\end{proposition}

An SI-unaware design, which ignores the floor, behaves as if $\bar\rho=1$ and changes over at \revold{$\alpha_{\rm r}=1/2$}, too early; the joint design keeps serving the users up to \revold{$\alpha_{\rm r}^\star>1/2$, e.g., $\alpha_{\rm r}^\star\approx0.60$} for $\epsilon^2\,\mathrm{INR}=-3$~dB. \revold{Equivalently, the floor multiplies the change-over odds $\alpha_{\rm r}^\star/(1-\alpha_{\rm r}^\star)$ by $1+\epsilon^2\,\mathrm{INR}\,(1+\varrho N_r)$; without the normalization, i.e., for $\alpha_{\rm c}\sum_k\gamma_k+\alpha_{\rm r}\gamma_q$, they would in addition be scaled by $\bar\Gamma_{\rm c}/\bar\gamma_q$, about $19$ and $22$~dB in Section~\ref{sec:simulations}, which moves the change-over to $\alpha_{\rm r}\approx0.99$.} With users that are only nearly orthogonal, the change-over is spread over a narrow range of \revold{$\alpha_{\rm r}$ around $\alpha_{\rm r}^\star$}. The same linearity drives the allocation among several targets (Section~\ref{sec:fairness}).

\section{Problem Formulation}
\label{sec:problem}

We jointly optimize the communication beamformers, the sensing covariance, the $Q$ radar combiners, and the delay-selective analog and digital cancellers. By Proposition~\ref{prop:support}, the support restriction is imposed as a constraint without loss of optimality. The weighted sum-SINR maximization problem is
 \begin{subequations}
\label{eq:P}
\begin{align}
\underset{\substack{\{\bw_k\},\,\bR_{\rm s},\,\{\bu_q\},\\\{\bC_{\rm A}[d]\},\,\{\bC_{\rm D}[d]\}}}{\operatorname{maximize}}\quad&{\revoldcolor\alpha_{\rm c}\sum_{k=1}^{K}\frac{\gamma_k}{\bar\Gamma_{\rm c}}+\frac{\alpha_{\rm r}}{Q}\sum_{q=1}^{Q}\frac{\gamma_q}{\bar\gamma_q}}
\label{eq:P_obj}\\
\operatorname{subject\ to}\quad&\tr(\bR_x)\le P_{\max},\label{eq:P_pow}\\
&\tr\big(\bR_{\rm A}^{\rm SI}\big)\le P_{\rm ADC}^{\max},\label{eq:P_adc}\\
&\textstyle\sum_{d}\|\bC_{\rm A}[d]\|_F^2\le P_{\rm A}^{\max},\label{eq:P_A}\\
&\textstyle\sum_{d}\|\bC_{\rm D}[d]\|_F^2\le P_{\rm D}^{\max},\label{eq:P_D}\\
&\bC_{\rm A}[d]=\bC_{\rm D}[d]=\mathbf{0},\ d\notin\Dsi,\label{eq:P_supp}\\
&\|\bu_q\|_2=1,\ q\in\mathcal{Q},\quad\bR_{\rm s}\succeq\mathbf{0}.\label{eq:P_rest}
\end{align}
\end{subequations}
\revold{The weights $\alpha_{\rm c},\alpha_{\rm r}\ge0$, $\alpha_{\rm c}+\alpha_{\rm r}=1$, set the priority of communication and sensing, and each function is normalized by its HD benchmark, $\bar\Gamma_{\rm c}$ of Section~\ref{sec:changeover} and $\bar\gamma_q$ of~\eqref{eq:hd_radar}. The normalization makes $\alpha_{\rm r}$ scale-free: without it, the sensing term would matter only for $\alpha_{\rm r}$ close to one (Section~\ref{sec:changeover}); with it, $\alpha_{\rm r}=0$ and $1$ give the communication-only and the sensing-only design, and the change-over lies at $\alpha_{\rm r}^\star$. All users share one weight, so the objective favors the strongest channels, as does $\bar\Gamma_{\rm c}$; per-user priorities, like the per-target weights of Section~\ref{sec:fairness}, only change the per-SINR weights of Section~\ref{sec:algorithm}.} The threshold $P_{\rm ADC}^{\max}$ in~\eqref{eq:P_adc} is the admissible SI power at the ADC inputs, summed over the $N_r$ branches; we set it to the level at which the quantization noise equals the thermal noise, $P_{\rm ADC}^{\max}=N_r\sigma_r^2\,10^{\mathrm{SQNR}/10}$, and below it the quantization noise is accounted for in~\eqref{eq:Ri}. The significantly weaker target and noise powers are neglected in~\eqref{eq:P_adc}. Constraints~\eqref{eq:P_A} and~\eqref{eq:P_D} impose tap-norm budgets, which are dimensionless (as $\|\bH_{\rm SI}[d]\|_F^2$) and are specified relative to $\sum_d\|\bHh_{\rm SI}[d]\|_F^2$ through $\xi_{\rm A}$ and $\xi_{\rm D}$.

Problem~\eqref{eq:P} couples beamforming and cancellation because $\bR_x$ determines the communication SINRs, the illumination of every target, the pre-ADC SI power, and the residual SI in every target bin, while the cancellers determine how much generated SI survives the analog and digital stages. It is non-convex: the objective is a sum of $K+Q$ fractional quadratic functions, and the transmit and cancellation variables are coupled through~\eqref{eq:RA} and~\eqref{eq:Rres}.

\section{Proposed Algorithm}
\label{sec:algorithm}

We employ the quadratic transform~\cite{shen2018fp1} together with block-coordinate optimization. For fixed auxiliary and dual variables, every primal block admits a closed-form update. \revold{The updates hold for the weighted sum-SINR $\mathcal{J}=\sum_k\alpha_k\gamma_k+\sum_q\alpha_q\gamma_q$ of Proposition~\ref{prop:support} with any nonnegative per-SINR weights;~\eqref{eq:P_obj} is the case $\alpha_k=\alpha_{\rm c}/\bar\Gamma_{\rm c}$ for every $k$ and $\alpha_q=\alpha_{\rm r}/(Q\bar\gamma_q)$.} The derivation assumes continuous, fully connected cancellation matrices over $\Dsi$; the effect of projecting the analog solution onto quantized hardware is evaluated in Section~\ref{sec:simulations}.

\subsection{Equivalent Transmit-Matrix Representation}

Since $\bR_{\rm s}\succeq\mathbf{0}$, write $\bR_{\rm s}=\bV\bV^H$ with $\bV\in\Cbb^{N_t\times N_s}$; $N_s=N_t$ represents any feasible sensing covariance. Define $\bW=[\bw_1,\ldots,\bw_K,\bV]=[\bq_1,\ldots,\bq_M]$, $M=K+N_s$, so that $\bR_x=\bW\bW^H=\sum_{m=1}^{M}\bq_m\bq_m^H$. For user $k$, $\gamma_k=|\bh_k^H\bq_k|^2/I_k$ with $I_k=\sum_{m\neq k}|\bh_k^H\bq_m|^2+\sigma_k^2$. Let $\bU_q=\bu_q\bu_q^H$, define the post-analog coupling matrix
\begin{equation}
\bG_{\rm A}=\sum_{d\in\Dsi}\bHh_{\rm A}^H[d]\bHh_{\rm A}[d]+N_r\sigma_\Delta^2\bI_{N_t},
\label{eq:GA}
\end{equation}
so that $\tr(\bR_{\rm A}^{\rm SI})=\sum_m\bq_m^H\bG_{\rm A}\bq_m$, and, for each target, the effective signature and target-bin SI matrix
\begin{align}
\ba_q&=\sqrt{\Np\sigma_q^2}\,\bH_q^H\bu_q,\label{eq:aq}\\
\bB_q&=\sum_{d\in\Dsi}\bHh_{\rm res}^H[d]\bU_q\bHh_{\rm res}[d]+\sigma_\Delta^2\bI_{N_t}+\varrho\,\bG_{\rm A}.\label{eq:Bq}
\end{align}
The radar SINR of target $q$ becomes
\begin{equation}
\gamma_q=\frac{\sum_{m=1}^{M}|\ba_q^H\bq_m|^2}{D_q},\qquad D_q=\sum_{m=1}^{M}\bq_m^H\bB_q\bq_m+\sigma_r^2.
\label{eq:sinr_q_compact}
\end{equation}

\subsection{Quadratic Transform}
\label{sec:qt}

For each user, introduce $y_k\in\Cbb$; the quadratic-transform identity gives
\begin{equation}
\gamma_k=\max_{y_k}\big[2\Re\{y_k^*\bh_k^H\bq_k\}-|y_k|^2I_k\big],
\label{eq:qt_k}
\end{equation}
maximized by $y_k=\bh_k^H\bq_k/I_k$. For each target, introduce $\bz_q\in\Cbb^{M}$; then
\begin{equation}
\gamma_q=\max_{\bz_q}\big[2\Re\{\bz_q^H\bW^H\ba_q\}-\|\bz_q\|_2^2D_q\big],
\label{eq:qt_q}
\end{equation}
maximized by $\bz_q=\bW^H\ba_q/D_q$. With all auxiliary variables fixed, the objective is a concave quadratic in $\bW$ and in each canceller ($D_q$ is a convex quadratic in each), which is what makes the closed-form updates possible.

\subsection{Transmit-Beamforming Update}

For stream $m$, let $\mathcal{K}_m=\mathcal{K}\setminus\{m\}$ for $m\le K$ and $\mathcal{K}_m=\mathcal{K}$ for $m>K$. The quadratic and linear coefficients are
\begin{align}
\bQ_m&=\sum_{q=1}^{Q}\alpha_q\|\bz_q\|_2^2\bB_q+\sum_{k\in\mathcal{K}_m}\alpha_k|y_k|^2\bh_k\bh_k^H,\label{eq:Qm}\\
\bb_m&=\sum_{q=1}^{Q}\alpha_qz_{q,m}^*\ba_q+\begin{cases}\alpha_my_m\bh_m,&m\le K,\\ \mathbf{0},&m>K,\end{cases}\label{eq:bm}
\end{align}
and the transmit subproblem is
\begin{subequations}
\label{eq:P_W}
\begin{align}
\underset{\{\bq_m\}}{\operatorname{maximize}}\quad&\sum_{m=1}^{M}\big[2\Re\{\bb_m^H\bq_m\}-\bq_m^H\bQ_m\bq_m\big]\\
\operatorname{subject\ to}\quad&\sum_{m}\|\bq_m\|_2^2\le P_{\max},\quad\sum_{m}\bq_m^H\bG_{\rm A}\bq_m\le P_{\rm ADC}^{\max}.
\end{align}
\end{subequations}
With dual variables $\eta,\mu\ge0$ for the power and ADC constraints, the \rev{Karush--Kuhn--Tucker (KKT)} conditions yield
\begin{equation}
\boxed{\;\bq_m=\big(\bQ_m+\eta\bI_{N_t}+\mu\bG_{\rm A}\big)^{-1}\bb_m\;}
\label{eq:W_update}
\end{equation}
(the Moore--Penrose inverse if singular). Since $\bQ_m$ takes only $K+1$ distinct values, $(\eta,\mu)$ is found by a nested bisection: for each $\mu$ the eigendecompositions of $\bQ_m+\mu\bG_{\rm A}$, $m\le K+1$, are computed once, after which each evaluation of the power constraint costs $\mathcal{O}(MN_t)$; the outer bisection is valid because, after minimization over $\eta$, the derivative of the dual function in $\mu$ is the slack of the pre-ADC constraint, which is nondecreasing in $\mu$. The first $K$ columns of $\bW$ are the beamformers and the remaining ones recover $\bR_{\rm s}=\bV\bV^H$.

\subsection{Radar-Combiner Update}

For fixed transmit and cancellation variables, the combiner of target $q$ maximizes $\bu_q^H\bR_{\rm tar,q}\bu_q/\bu_q^H\bR_{\rm int}\bu_q$ with $\bR_{\rm tar,q}=\Np\bR_q$ and $\bR_{\rm int}=\bR_{\rm i}$ from~\eqref{eq:Ri}, whose solution is the dominant generalized eigenvector of $(\bR_{\rm tar,q},\bR_{\rm int})$. For the rank-one target channel, $\bR_{\rm tar,q}\propto\ba_r(\theta_q)\ba_r^H(\theta_q)$ and the solution reduces exactly to the minimum-variance distortionless-response form
\begin{equation}
\boxed{\;\bu_q=\frac{\bR_{\rm int}^{-1}\ba_r(\theta_q)}{\|\bR_{\rm int}^{-1}\ba_r(\theta_q)\|_2}\;}
\label{eq:u_update}
\end{equation}
Since $\bR_{\rm int}$ is common to all targets, the $Q$ combiners are obtained from one factorization and $Q$ triangular solves.

\subsection{Unified Canceller Update}

Both cancellers, as well as the minimizer in Lemma~\ref{lem:budget}, solve a problem of the following form.

\begin{lemma}[Generalized Sylvester update]
\label{lem:sylvester}
Let $\bL\succeq\mathbf{0}$ with eigendecomposition $\bL=\mathbf{E}\bLambda\mathbf{E}^H$, let $\bR_x=\mathbf{F}\bSigma\mathbf{F}^H$ with $\bSigma=\diag(p_1,\ldots,p_{N_t})$, and let $\{\bT_d\}_{d\in\Dsi}$ be given. The problem
\begin{equation}
\underset{\{\bC[d]\}}{\operatorname{minimize}}\ \sum_{d\in\Dsi}\Big[\tr\big(\bL\bC[d]\bR_x\bC^H[d]\big)-2\Re\tr\big(\bT_d\bR_x\bC^H[d]\big)\Big]
\label{eq:sylv_problem}
\end{equation}
subject to $\sum_d\|\bC[d]\|_F^2\le P_{\rm C}$ is solved by $\bC[d]=\mathbf{E}\widetilde{\bC}_d\mathbf{F}^H$ with
\begin{equation}
\boxed{\;\big[\widetilde{\bC}_d\big]_{i,j}=\frac{[\widetilde{\bT}_d]_{i,j}\,p_j}{[\bLambda]_{i,i}\,p_j+\vartheta}\;}\qquad
\widetilde{\bT}_d=\mathbf{E}^H\bT_d\mathbf{F},
\label{eq:sylv_solution}
\end{equation}
where entries with a vanishing denominator are set to zero (minimum-norm solution; with the regularization of Section~\ref{sec:tiebreak}, no denominator vanishes), and $\vartheta\ge0$ is zero if the unconstrained solution satisfies the budget and otherwise the unique root of $\sum_d\|\widetilde{\bC}_d\|_F^2=P_{\rm C}$, found by bisection.
\end{lemma}

The digital and analog updates follow by identifying $(\bL,\bT_d)$. Because the cancellers enter the quadratic-transformed objective only through $-\sum_q\alpha_q\|\bz_q\|_2^2D_q$, the relevant weighting is
\begin{equation}
\bL_{\rm D}=\sum_{q=1}^{Q}w_q\bU_q,\qquad w_q=\alpha_q\|\bz_q\|_2^2,\qquad \bar w=\sum_{q}w_q,
\label{eq:LD}
\end{equation}
with $\bz_q$ recomputed from~\eqref{eq:qt_q} at the current iterate.

\subsubsection{Digital canceller}
For fixed $\bW$, $\{\bu_q\}$, and analog canceller, the digital canceller affects only the target-bin residual SI, and minimizing $\sum_d\tr(\bL_{\rm D}\bHh_{\rm res}[d]\bR_x\bHh_{\rm res}^H[d])$ subject to~\eqref{eq:P_D} is~\eqref{eq:sylv_problem} with $\bL=\bL_{\rm D}$ and $\bT_d=\bL_{\rm D}\bHh_{\rm A}[d]$, evaluated with the regularization of Section~\ref{sec:tiebreak}. For a single target, $\bL_{\rm D}=w\bU$\revold{, so that $\widetilde{\bL}_{\rm D}=w(\bU+\xi\bI_{N_r})$, and~\eqref{eq:sylv_solution} with the regularization of Section~\ref{sec:tiebreak} becomes}
\begin{equation}
\revoldcolor\begin{aligned}
\bC_{\rm D}[d]={}&\bU\bHh_{\rm A}[d]\widetilde{\bR}_x(\widetilde{\bR}_x+\vartheta_{\rm D}\bI_{N_t})^{-1}\\
&+(\bI_{N_r}-\bU)\bHh_{\rm A}[d]\widetilde{\bR}_x(\widetilde{\bR}_x+\varsigma\bI_{N_t})^{-1},
\end{aligned}
\label{eq:CD_single}
\end{equation}
\revold{where $\widetilde{\bR}_x=\bR_x+\xi\tr(\bR_x)\bI_{N_t}$, $\vartheta_{\rm D}=\vartheta/(w(1+\xi))$, and $\varsigma=\vartheta_{\rm D}(1+\xi)/\xi$, with $\vartheta$ the multiplier of~\eqref{eq:P_D}. As $\xi\to0$, either the budget binds along $\bu$, so that $\vartheta_{\rm D}>0$ and $\varsigma\to\infty$, or $\vartheta_{\rm D}\to0$ and $\varsigma$ remains finite, set by the budget left by the first term. The first term is the minimizer along $\bu$, unique on the range of $\bR_x$; the second does not affect the objective and spends the remaining budget on the SI outside $\bu$, including the transmit directions that currently carry no signal.}

\subsubsection{Analog canceller}
For fixed $\bW$, $\{\bu_q\}$, and digital canceller, the analog canceller reduces the target-bin residual SI and the quantization noise, and satisfies the pre-ADC constraint:
\begin{subequations}
\label{eq:P_A_sub}
\begin{align}
\underset{\{\bC_{\rm A}[d]\}}{\operatorname{minimize}}\quad&\sum_{d}\tr\big(\bL_{\rm D}\bHh_{\rm res}[d]\bR_x\bHh_{\rm res}^H[d]\big)+\bar w\varrho\,\Pi_{\rm A}\\
\operatorname{subject\ to}\quad&\Pi_{\rm A}\le\widetilde{P}_{\rm ADC},\quad\eqref{eq:P_A},
\end{align}
\end{subequations}
where $\Pi_{\rm A}=\sum_d\tr(\bHh_{\rm A}[d]\bR_x\bHh_{\rm A}^H[d])$ and $\widetilde{P}_{\rm ADC}=P_{\rm ADC}^{\max}-N_r\sigma_\Delta^2\tr(\bR_x)$. With dual variable $\nu\ge0$ for the pre-ADC constraint, this is~\eqref{eq:sylv_problem} with
\begin{align}
\bL&=\widetilde{\bL}_{\rm D}+(\nu+\bar w\varrho)\bI_{N_r},\nonumber\\
\bT_d&=\widetilde{\bL}_{\rm D}\big(\bHh_{\rm SI}[d]-\bC_{\rm D}[d]\big)+(\nu+\bar w\varrho)\bHh_{\rm SI}[d],
\label{eq:analog_LT}
\end{align}
where $\widetilde{\bL}_{\rm D}$ is the regularized weighting of Section~\ref{sec:tiebreak}, and $\nu$ is found by an outer bisection on the pre-ADC constraint, which is valid by the same convexity argument as for $\mu$. For a single target, $\mathbf{E}$ is any unitary matrix with first column $\bu$ and, up to the regularization, $\bLambda=\diag(w+\nu+\varrho w,\nu+\varrho w,\ldots)$, so no eigendecomposition is needed. Lemma~\ref{lem:budget} is the special case $\bL=\bI_{N_r}$, $\bT_d=\bHh_{\rm SI}[d]$, i.e., the limit $\nu\to\infty$.

\subsubsection{Tie-breaking among minimizers}
\label{sec:tiebreak}
\revold{The digital subproblem is flat along the null spaces of $\bR_x$ and of $\bL_{\rm D}$, and the analog one along the null space of $\bR_x$}, so their minimizers are not unique, and the minimum-norm choice reconstructs nothing in the unused transmit and receive directions; the subsequent transmit and combiner updates would then see a large residual SI outside the current subspaces and remain locked to them. We therefore evaluate~\eqref{eq:sylv_solution} with $\bR_x+\xi\tr(\bR_x)\bI_{N_t}$ in place of $\bR_x$ and $\widetilde{\bL}_{\rm D}=\bL_{\rm D}+\xi\tr(\bL_{\rm D})\bI_{N_r}$ in place of $\bL_{\rm D}$, $\xi=10^{-6}$, \emph{both} in $\bL$ and in $\bT_d$ (with $\bT_d=\bL_{\rm D}\bHh_{\rm A}[d]$ alone, $\widetilde{\bT}_d$ would vanish outside the range of $\bL_{\rm D}$). Among the minimizers, this selects the cancellers that also reconstruct the SI in the unused directions, as far as the budget allows;~\eqref{eq:CD_single} is its single-target \revold{form}. As $\xi\to0$, the selected cancellers remain minimizers of the original subproblems, and the choice is deterministic rather than left to floating-point noise.

\subsection{Fairness Among Targets}
\label{sec:fairness}

The objective~\eqref{eq:P_obj} is linear in the radar SINRs, and each $\gamma_q$ is linear in the power allocated to target $q$ once the SI is cancelled, so, as in Proposition~\ref{prop:changeover}, a weighted sum concentrates all sensing power on the target with the largest weighted gain unless the weights are exactly balanced. Max-min fairness is obtained by \revold{weighting the targets individually within the sensing term, $\alpha_q=\alpha_{\rm r}\omega_q/(Q\bar\gamma_q)$ with $\omega_q\ge0$ and $\sum_q\omega_q=Q$, which leaves $\alpha_{\rm r}$ unchanged. Starting from $\omega_q=1$}, after each outer step we set
\begin{equation}
\revoldcolor\omega_q\leftarrow\omega_q\Big(\frac{\gamma_{\rm g}}{\gamma_q}\Big)^{\kappa},\qquad
\gamma_{\rm g}=\Big(\prod_{q}\gamma_q\Big)^{1/Q},
\label{eq:fair_update}
\end{equation}
\revold{renormalize to $\sum_q\omega_q=Q$}, and run a few iterations of Algorithm~\ref{alg:joint} from the previous iterate. The damping $\kappa\in(0,1)$ and the warm start are essential: with $\kappa=1$ and an exact inner solve the iteration jumps between extreme points, whereas with $\kappa=0.3$ and three inner iterations the SINRs equalize within a few outer steps. Since every iterate is feasible, the one with the largest minimum SINR is retained. The procedure is a heuristic, with no optimality claim beyond feasibility and monotone inner iterations, and is used in Section~\ref{sec:simulations} for $Q>1$.

\subsection{Algorithm, Convergence, and Complexity}
\label{sec:complexity}

The complete procedure is summarized in Algorithm~\ref{alg:joint}. \revold{Because~\eqref{eq:P_obj} is linear in the SINRs, \rev{the point at which the algorithm settles depends on its start and typically lies near the communication or the sensing face of the trade-off region}, and a stream with zero power remains zero ($\bq_m=\mathbf{0}$ gives $y_m=0$, $z_{q,m}=0$, and $\bb_m=\mathbf{0}$ in~\eqref{eq:bm}). Algorithm~\ref{alg:joint} is therefore run from three starting points, and the best output is retained. The first is the SI-free design at the same weights (steps 3--6 with $\bB_q=\mathbf{0}$ and without~\eqref{eq:P_adc}), followed by the canceller updates for the resulting $\bR_x$ and a uniform power back-off if~\eqref{eq:P_adc} cannot be met, i.e., the sequential design of Section~\ref{sec:setup}; since it is feasible, the joint design never does worse. The other two are cold starts with maximum-ratio user beams carrying $50\%$ and $95\%$ of the power, an isotropic sensing signal carrying the rest, the SI-channel estimate scaled to the tap budget as analog canceller, and the same back-off.}

\begin{algorithm}[t]
\caption{Joint Beamforming and Delay-Selective SI Cancellation for $K$ Users and $Q$ Targets}
\label{alg:joint}
\begin{algorithmic}[1]
\Require Channels $\{\bh_k\}$, $\{\bH_q\}$, estimates $\{\bHh_{\rm SI}[d]\}$, weights, tolerance $\epsilon_{\rm tol}$, $T_{\max}$
\Ensure $\bW$, $\{\bu_q\}$, $\{\bC_{\rm A}[d]\}$, $\{\bC_{\rm D}[d]\}$
\State \revold{Initialize with the SI-free design or one of the two cold starts (run all three, keep the best output)}; $r\leftarrow0$
\Repeat
\State Form $\bR_x$, $\bHh_{\rm A}[d]$, $\bHh_{\rm res}[d]$, $\bG_{\rm A}$, $\{\ba_q,\bB_q\}$
\State Update $\{y_k\}$ and $\{\bz_q\}$ (Section~\ref{sec:qt})
\State Update $\bW$ by~\eqref{eq:W_update} with nested bisection on $(\eta,\mu)$
\State Recompute $\bR_x$; update $\{\bu_q\}$ by~\eqref{eq:u_update}
\State Recompute $\{\bz_q\}$, form $\widetilde{\bL}_{\rm D}$; update $\{\bC_{\rm A}[d]\}$ by Lemma~\ref{lem:sylvester} with~\eqref{eq:analog_LT} and bisection on $\nu$
\State Update $\{\bC_{\rm D}[d]\}$ by Lemma~\ref{lem:sylvester} with $\bL=\widetilde{\bL}_{\rm D}$, $\bT_d=\widetilde{\bL}_{\rm D}\bHh_{\rm A}[d]$
\State Evaluate~\eqref{eq:P_obj}; $r\leftarrow r+1$
\Until{relative objective change $\le\epsilon_{\rm tol}$ or $r\ge T_{\max}$}
\end{algorithmic}
\end{algorithm}

\begin{proposition}[Monotone convergence]
\label{prop:conv}
\revold{With the canceller updates evaluated in the limit $\xi\to0$ of Section~\ref{sec:tiebreak}, the} sequence of objective values~\eqref{eq:P_obj} generated by Algorithm~\ref{alg:joint} is nondecreasing and converges\revold{; for $\xi>0$, each canceller update can lower~\eqref{eq:P_obj} by at most $\mathcal{O}(\xi)$}.
\end{proposition}
\begin{IEEEproof}
Every step keeps the iterate feasible. The auxiliary update~\eqref{eq:qt_k}--\eqref{eq:qt_q} makes the transformed objective equal to~\eqref{eq:P_obj}, which it lower-bounds for any auxiliary values. The update~\eqref{eq:W_update} maximizes the transformed objective over $\bW$ (a concave quadratic program with the convex constraints of~\eqref{eq:P_W}); each $\bu_q$ by~\eqref{eq:u_update} maximizes $\gamma_q$ itself, and the communication SINRs do not depend on it; after the auxiliary variables are recomputed, each canceller update by Lemma~\ref{lem:sylvester}, with the pre-ADC constraint enforced through $\nu$, maximizes the transformed objective over that canceller, exactly for $\xi\to0$ and up to $\mathcal{O}(\xi)$ otherwise. Hence~\eqref{eq:P_obj} is nondecreasing. It is bounded above by $\sum_k\alpha_k\gamma_k^{\rm su}+\sum_q\alpha_q\gamma_q^\star$, where $\gamma_k^{\rm su}=P_{\max}\|\bh_k\|_2^2/\sigma_k^2$ is the single-user SINR and $\gamma_q^\star$ is the ceiling~\eqref{eq:ceiling}, hence it converges.
\end{IEEEproof}

\rev{Proposition~\ref{prop:conv} guarantees the convergence of the objective values, not that the iterates converge to a stationary point of~\eqref{eq:P}: the block-coordinate results of~\cite{razaviyayn2013unified,bertsekas1999nonlinear} require constraints that separate across the blocks, whereas~\eqref{eq:P_adc} couples $\bW$ and the analog canceller. This is one more reason to run Algorithm~\ref{alg:joint} from several starting points.}

\begin{table}[t]
\centering
\caption{Per-Iteration Complexity of Algorithm~\ref{alg:joint}}
\label{tab:complexity}
\footnotesize
\begin{tabular}{@{}ll@{}}
\toprule
Block & Complexity \\
\midrule
Auxiliary variables & $\mathcal{O}\big((K+Q)MN_t\big)$ \\
Transmit update~\eqref{eq:W_update} & $\mathcal{O}\big(I_\mu[(K+1)N_t^3+I_\eta MN_t]\big)$ \\
Combiners~\eqref{eq:u_update} & \revold{$\mathcal{O}\big(N_r^3+QN_r^2+D_{\rm SI}N_rN_t(N_r+N_t)\big)$} \\
Analog canceller & $\mathcal{O}\big(I_\nu[N_r^3+D_{\rm SI}N_rN_t(N_r+N_t)+I_\vartheta D_{\rm SI}N_rN_t]\big)$ \\
Digital canceller & $\mathcal{O}\big(N_r^3+N_t^3+D_{\rm SI}N_rN_t(N_r+N_t)+I_\vartheta D_{\rm SI}N_rN_t\big)$ \\
\bottomrule
\end{tabular}
\end{table}

Table~\ref{tab:complexity} lists the per-iteration cost, where $I_\eta$, $I_\vartheta$, $I_\mu$, and $I_\nu$ are the numbers of bisection steps (typically $30$--$60$). The complexity is cubic in the array size and linear in $K$, $Q$, and $D_{\rm SI}$; no interior-point solver is involved. For $N_t=N_r=16$, $K=3$, $Q=1$, one iteration takes about \revold{$43$}~ms in unoptimized interpreted code on one core of a cloud virtual machine, and convergence to a relative tolerance of $10^{-5}$ takes between \revold{$2$ and $43$} iterations depending on the operating point (Section~\ref{sec:simulations}).

\section{Simulation Results}
\label{sec:simulations}
\begingroup\revoldcolor  

\subsection{Setup and Benchmarks}
\label{sec:setup}

We consider a very-low-LEO satellite at $h=300$~km, operating at $20$~GHz with $1/T_s=20$~MBd and a $43$~dBm RF budget. It serves $K=3$ users and senses $Q=1$ large vessel at $60^\circ$ elevation with $N_t=N_r=N\in\{8,16\}$ ports. Users and targets are uniformly distributed in the $\pm4^\circ$ field of view with line-of-sight channels, and the fractional target delay is uniform, so that the straddle factor of Section~\ref{sec:rc} is included in $\bar\gamma_q$. The SI channel~\eqref{eq:si_taps} comprises the near-field path~\eqref{eq:nearfield} between parallel arrays $0.5$~m apart and two Rayleigh platform reflections, $10$ and $20$~dB weaker, at $15$ and $35$~ns, sampled by the raised-cosine response (roll-off $0.35$) on $D_{\rm pre}=D_{\rm post}=10$. Table~\ref{tab:params} lists all parameters. All designs are evaluated with~\eqref{eq:sinr_k} and~\eqref{eq:sinr_q}, i.e., with the residual SI, the non-reconstructible SI, and the quantization noise of the actual design. Results are averaged in the linear domain over $15$ channel realizations, common to all figures except Fig.~\ref{fig:ls}, and then converted to decibels.

\begin{table}[t]
\centering
\caption{Simulation Parameters and Link Budget}
\label{tab:params}
\footnotesize
\setlength{\tabcolsep}{3pt}
\begin{tabularx}{\columnwidth}{@{}lX@{}}
\toprule
Geometry & $h=300$~km, elevation $60^\circ$, $R_q=343.9$~km, $d_q=45\,879$\\
Waveform & $f_c=20$~GHz, $1/T_s=20$~MBd, RRC roll-off $0.35$\\
Transmitter & $P_{\max}=43$~dBm, $N_t=N_r=N\in\{8,16\}$ ports of $G_e=25$~dBi, spacing $7\lambda_c$\\
Users & $K=3$, $G_u=20$~dBi, $T_u=400$~K, $L_a=2$~dB; $\bar\Gamma_{\rm c}=25.4$/$28.4$~dB ($N=8$/$16$)\\
Sensing & $T_{\rm sys}=500$~K, RCS $\sigma_q^2=40$~dBsm, $\Tcpi=5$~ms, $\Np=10^5$\\
SI channel & $\kappa_{\rm SI}=-40$~dB, reflections $-10$/$-20$~dB at $15$/$35$~ns; $\mathrm{INR}=48.4$/$46.9$~dB\\
Impairments & $\epsilon^2=-50$~dB; $b=8$ bits, PAPR $10$~dB (SQNR $39.9$~dB)\\
Budgets & $\xi_{\rm A}=0.1$ ($0.02$ in Fig.~\ref{fig:adc}), $\xi_{\rm D}=1$\\
\midrule
Sensing SNR & $\bar\gamma_q=3.1$/$9.2$~dB, i.e., $-55.9$/$-52.9$~dB per sample and receive branch\\
Closed forms & $\gamma_q^\star=0.85$/$7.44$~dB~\eqref{eq:ceiling}; $\alpha_{\rm r}^\star=0.63$/$0.60$~\eqref{eq:alpha_star}\\
\bottomrule
\end{tabularx}
\end{table}

We compare with four references. \emph{HD, communication only:} the sum-SINR $\bar\Gamma_{\rm c}$ with the whole power on the downlink and no SI. \emph{HD, sensing only:} $\bar\gamma_q$ of~\eqref{eq:hd_radar}, and $(\sum_q1/\bar\gamma_q)^{-1}$ for $Q$ targets with equal SINRs. Both are single-function bounds that no system attains simultaneously. \emph{SI-unaware:} the SI-free design at the same weights, i.e., steps 3--6 of Algorithm~\ref{alg:joint} with $\bB_q=\mathbf{0}$ and without~\eqref{eq:P_adc}, deployed without cancellation, with $\bu_q=\ba_r/\sqrt{N_r}$ and no back-off. \rev{It violates~\eqref{eq:P_adc} and is therefore not a feasible design; it serves only as a reference for the cost of ignoring the SI.} \emph{Sequential:} the same SI-free beamformers, backed off uniformly until~\eqref{eq:P_adc} can be met with the canceller of Lemma~\ref{lem:budget}, followed by the combiner and both cancellers optimized for the resulting $\bR_x$. It handles the SI at the receiver only, is the first starting point of Algorithm~\ref{alg:joint}, and isolates the gain of designing the transmitter jointly with the cancellers\rev{; it is the relevant feasible benchmark}.

\begin{figure}[t]\centering\includegraphics[width=\columnwidth]{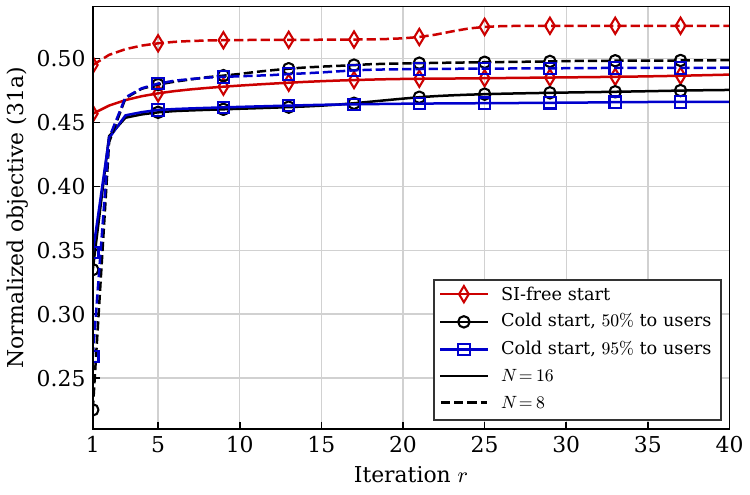}
\caption{Normalized objective~\eqref{eq:P_obj} versus iteration at $\alpha_{\rm r}=0.55$ for the three starting points of Algorithm~\ref{alg:joint}.}\label{fig:conv}\end{figure}

\subsection{Convergence}

Fig.~\ref{fig:conv} shows the objective at $\alpha_{\rm r}=0.55$, between the change-over weights $1/2$ of the SI-free design and $\alpha_{\rm r}^\star$. Every run is nondecreasing (Proposition~\ref{prop:conv}); the largest relative decrease over all runs of this section is below $10^{-9}$, i.e., at the numerical precision. All starts come within $10^{-3}$ of their final value after $9$ to $25$ iterations, but \rev{their final values differ}: $0.487$ ($N=16$) and $0.526$ ($N=8$) for the SI-free start, and $0.475$/$0.499$ and $0.466$/$0.493$ for the cold starts with $50\%$ and $95\%$ of the power on the users. Over the sweep of Fig.~\ref{fig:trade}, a cold start yields the best output at $11\%$ ($N=16$) and $16\%$ ($N=8$) of the points, all with $\alpha_{\rm r}\le0.65$, so all three starts are run; one iteration takes $43$~ms ($N=16$) and $12$~ms ($N=8$).

\begin{figure}[t]\centering\includegraphics[width=\columnwidth]{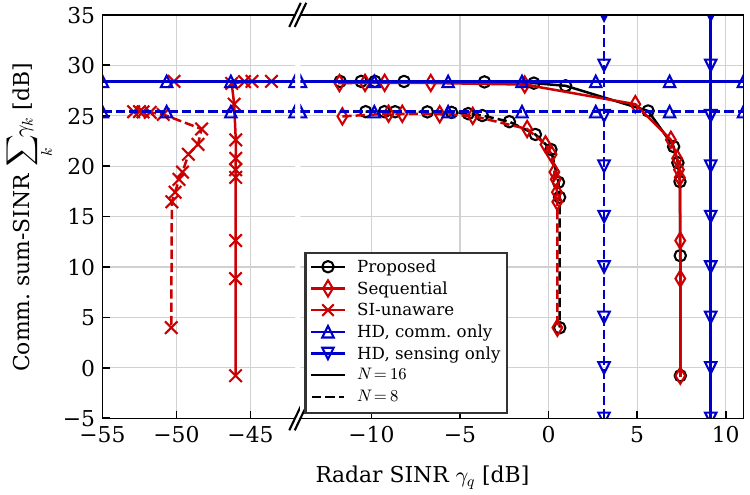}
\caption{Communication--sensing trade-off traced by $\alpha_{\rm r}\in[0,1]$; the radar-SINR axis is interrupted between $-42$ and $-14$~dB\rev{; the SI-unaware reference violates~\eqref{eq:P_adc}}.}\label{fig:trade}\end{figure}

\subsection{Communication--Sensing Trade-Off}

Fig.~\ref{fig:trade} maps the trade-off traced by $\alpha_{\rm r}\in[0,1]$. At $\alpha_{\rm r}=0$, the proposed design attains $\bar\Gamma_{\rm c}=28.4$~dB ($N=16$) and $25.4$~dB ($N=8$): FD sensing costs nothing on the downlink (Lemma~\ref{lem:budget}). The sequential design must back off in the $2$ ($N=16$) and $3$ ($N=8$) realizations in which the tap budget does not remove the SI of its beams, by $1.7$ and $3.2$~dB, and loses $0.2$ and $0.5$~dB on average. At $\alpha_{\rm r}=1$, the radar SINR equals the ceiling $\gamma_q^\star=7.44$~dB within $0.01$~dB for $N=16$, $1.7$~dB below the HD sensing SNR, as~\eqref{eq:ceiling} predicts for $\epsilon^2\,\mathrm{INR}=-3.1$~dB; for $N=8$ it is $0.63$~dB, $0.2$~dB below the ceiling, because $\xi_{\rm A}=0.1$ meets the condition of Proposition~\ref{prop:ceiling} in only $40\%$ of the realizations ($100\%$ for $N=16$, Corollary~\ref{cor:budget}).

In between, the proposed design keeps the downlink within $0.2$~dB of $\bar\Gamma_{\rm c}$ up to $\alpha_{\rm r}=0.5$ while its radar SINR rises from $-11.8$ to $-0.8$~dB ($N=16$), and it changes over between $\alpha_{\rm r}=0.55$ and $0.7$, around $\alpha_{\rm r}^\star$. The sequential design inherits the SI-free beams, which change over at $\alpha_{\rm r}=1/2$ (Proposition~\ref{prop:changeover}). Its points lie on the curve of the proposed design, within $0.3$~dB, but at the wrong weights: at $\alpha_{\rm r}=0.55$ and $N=16$, it delivers $6.9$~dB of radar SINR at $22.6$~dB of downlink, the point that the proposed design selects only for $\alpha_{\rm r}$ between $0.6$ and $0.65$, whereas the proposed design keeps $28.0$~dB at $1.0$~dB and a $10\%$ larger objective ($0.497$ against $0.446$). Above $\alpha_{\rm r}^\star$, the SI floor lowers the value of sensing power, so the proposed design gives part of the power back to the users at almost no cost in radar SINR and keeps $1.5$, $5.8$, and $2.3$~dB more downlink at $\alpha_{\rm r}=0.7$, $0.8$, and $0.9$. Overall, its objective exceeds that of the sequential design by up to $11\%$ ($N=16$) and $13\%$ ($N=8$). With nominal hardware, the known SI is removed completely and~\eqref{eq:P_adc} is inactive, so the problem nearly decouples and the joint design adds mainly the correct change-over. \rev{The infeasible SI-unaware reference, whose downlink equals that of the SI-free design, overloads the ADCs by $6.6$--$8.9$~dB and attains a radar SINR between $-52.9$ and $-43.6$~dB at every weight: ignoring the SI is not an option, and the comparisons below focus on the sequential design.}

\subsection{Confirmation of the Closed-Form Results}

\begin{figure}[t]\centering\includegraphics[width=\columnwidth]{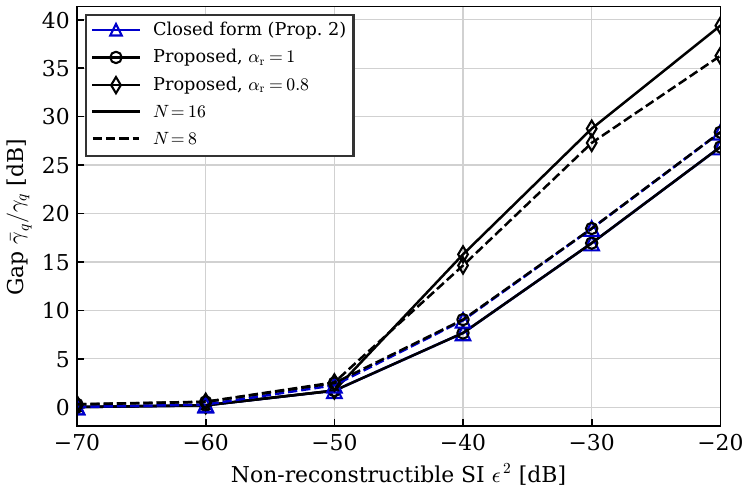}
\caption{Gap between the HD sensing SNR and the radar SINR of the proposed design versus $\epsilon^2$, with the closed form of Proposition~\ref{prop:ceiling}.}\label{fig:eps}\end{figure}

\emph{Ceiling (Proposition~\ref{prop:ceiling}).} In Fig.~\ref{fig:eps}, at $\alpha_{\rm r}=1$ and $N=16$, the simulated gap to the HD sensing SNR coincides with the closed form within $0.01$~dB for $\epsilon^2$ from $-70$ to $-20$~dB, which confirms~\eqref{eq:Rres} and the achievability part of the proposition; for $N=8$ it exceeds the closed form by $0.27$~dB at small $\epsilon^2$, the signature of $\xi_{\rm A}=0.1<1/8$ (Corollary~\ref{cor:budget}). Once $\epsilon^2\,\mathrm{INR}$ exceeds one, the non-reconstructible SI sets the sensing performance: above $-40$~dB, every $10$~dB of $\epsilon^2$ costs $9.3$--$9.9$~dB of radar SINR. At $\alpha_{\rm r}=0.8$, the gap jumps to $15.8$~dB ($N=16$) and $14.6$~dB ($N=8$) at $\epsilon^2=-40$~dB, where the design switches to the users: by~\eqref{eq:alpha_star}, $\alpha_{\rm r}^\star$ grows from $0.60$ to $0.85$ ($0.63$ to $0.89$ for $N=8$), so that $\alpha_{\rm r}=0.8$ falls below the change-over weight, as Proposition~\ref{prop:changeover} predicts.

\begin{figure}[t]\centering\includegraphics[width=\columnwidth]{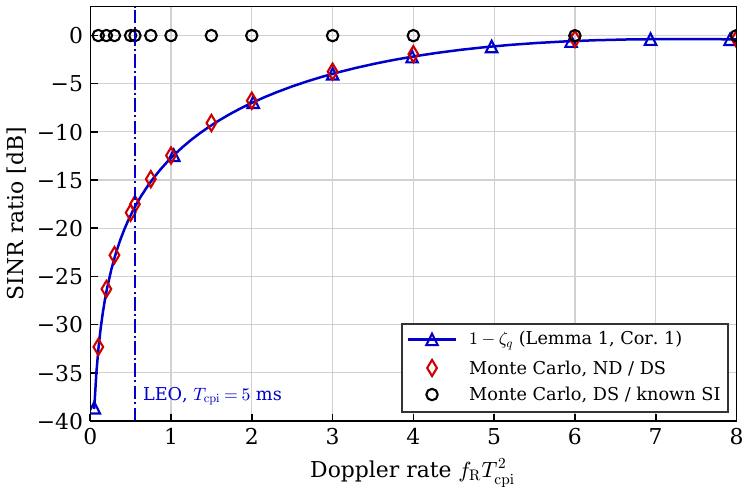}
\caption{LS cancellers estimated on the CPI at the zero-Doppler crossing ($4\times4$, $\Np=4000$, $\mathrm{INR}=40$~dB): SINR of the \rev{non-delay-selective (ND)} relative to the \rev{delay-selective (DS)} canceller, and of the DS canceller relative to the known-SI receiver.}\label{fig:ls}\end{figure}

\emph{Echo absorption (Lemma~\ref{lem:nd}).} Fig.~\ref{fig:ls} reports a time-domain Monte Carlo in which LS cancellers are estimated from $\Np=4000$ post-analog samples of a $4\times4$ system with $\mathrm{INR}=40$~dB and an echo $10$~dB below the noise per sample, and applied to the same block. The \rev{DS} canceller recovers the SINR of the known-SI receiver within $0.06$~dB, as stated in Lemma~\ref{lem:nd}(a), whereas without cancellation the SINR is $39.4$~dB lower. The \rev{ND} canceller, whose support contains $d_q$, follows $1-\zeta_q$ of~\eqref{eq:zeta_chirp} within $0.4$~dB; the same holds for a Doppler offset with~\eqref{eq:zeta_lin}, $f_{\rm D}\Tcpi\in[0.05,2]$. At the LEO operating point, $f_{\rm R}\Tcpi^2=0.55$, the loss is $17.5$~dB, against $17.8$~dB in closed form, and by~\eqref{eq:nd_loss} it falls below $1$~dB only for $f_{\rm D}\Tcpi\ge0.65$ or $f_{\rm R}\Tcpi^2\ge5.2$, when the echo decorrelates within the CPI.

\begin{figure}[t]\centering\includegraphics[width=\columnwidth]{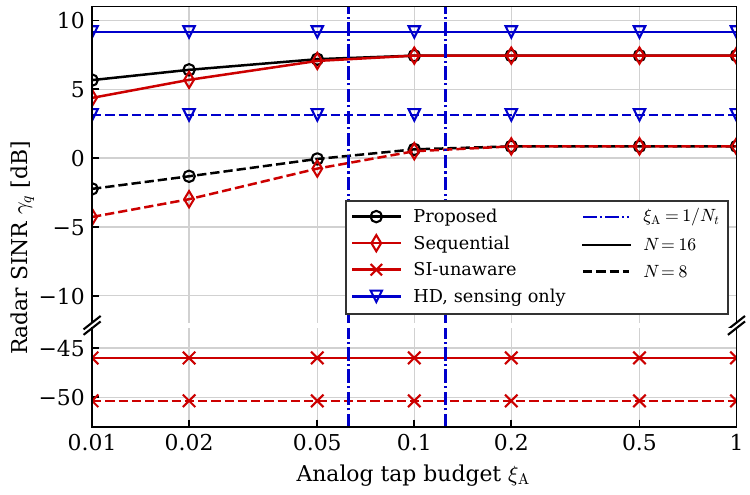}
\caption{Radar SINR versus the analog tap budget at $\alpha_{\rm r}=1$; the vertical axis is interrupted\rev{; the SI-unaware reference violates~\eqref{eq:P_adc}}.}\label{fig:tap}\end{figure}

\emph{Tap budget (Lemma~\ref{lem:budget}, Corollary~\ref{cor:budget}).} In Fig.~\ref{fig:tap}, the proposed design reaches the ceiling at $\xi_{\rm A}=0.1$ for $N=16$ and $0.2$ for $N=8$, the first grid points above $1/N_t$, as Corollary~\ref{cor:budget} predicts for a rank-one sensing covariance; over $N\in\{8,12,16,24,32\}$, its gap to the ceiling at $\xi_{\rm A}=0.05$ decreases as $0.92$, $0.47$, $0.26$, $0.14$, and $0.08$~dB. Below the threshold, it steers the beam away from the strongly coupled eigendirections and loses $1.8$ ($N=16$) and $3.1$~dB ($N=8$) with respect to the ceiling at $\xi_{\rm A}=0.01$, whereas the sequential design, which keeps the SI-free beam, must back off by $2.0$ and $4.4$~dB and loses $3.1$ and $5.1$~dB. Where the budget binds, the pre-ADC SI of the proposed design lies within $0.1$--$1.1$~dB of $\Pi_{\rm A}^{\min}$ of~\eqref{eq:pimin}; above $1/N_t$, $\Pi_{\rm A}^{\min}=0$ and~\eqref{eq:P_adc} is inactive. The \rev{infeasible} SI-unaware \rev{reference} stays at $-46.0$ and $-50.4$~dB for every budget, with $6.6$ and $8.0$~dB more SI at the ADCs than $P_{\rm ADC}^{\max}$.

\begin{figure}[t]\centering\includegraphics[width=\columnwidth]{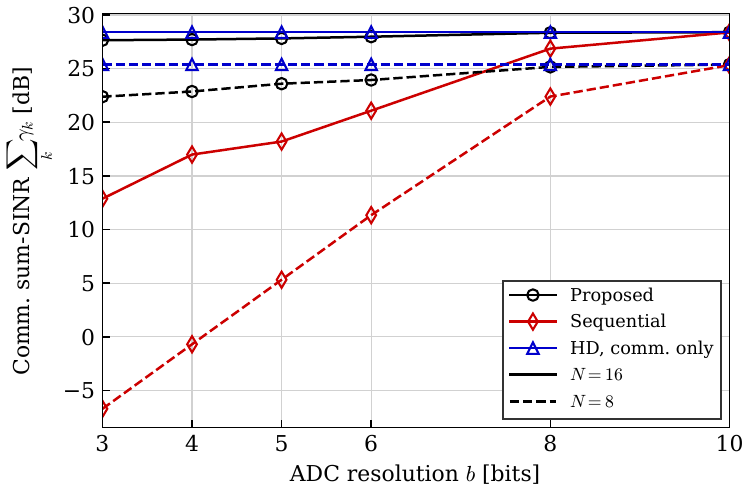}
\caption{Communication sum-SINR versus ADC resolution at $\alpha_{\rm r}=0.3$ with a hardware-limited analog canceller ($\xi_{\rm A}=0.02$).}\label{fig:adc}\end{figure}

\subsection{ADC Resolution and Hardware-Constrained Cancellation}

With $\xi_{\rm A}=0.1$ and $N=16$, the known SI is removed before the ADCs and the resolution hardly matters, so Fig.~\ref{fig:adc} uses $\xi_{\rm A}=0.02$, for which~\eqref{eq:P_adc} is active. At $\alpha_{\rm r}=0.3$, the proposed design keeps the downlink within $0.8$~dB of $\bar\Gamma_{\rm c}$ down to $3$ bits for $N=16$ ($27.6$ against $28.4$~dB) and within $3.0$~dB for $N=8$, operating on average at $89$--$100\%$ of $P_{\rm ADC}^{\max}$ for $b\le6$. The sequential design must back off by $2.2$ ($N=16$) and $3.6$~dB ($N=8$) already with $8$-bit ADCs and by $23$ and $27$~dB with $4$-bit ADCs, where its sum-SINR falls to $17.0$ and $-0.7$~dB. At $\alpha_{\rm r}=1$, the radar SINR of the proposed design remains within $3.0$~dB of the ceiling at the same resolution for $N=16$ and decreases to $-5.4$~dB for $N=8$, whereas that of the sequential design falls to $-12.5$ and $-24.9$~dB with $4$-bit ADCs. With $4$-bit ADCs, the joint design thus retains $10.7$~dB ($N=16$) and $23.6$~dB ($N=8$) more downlink and $17.3$ and $19.5$~dB more radar SINR than the sequential design. The gain comes from the multiplier $\mu$ in~\eqref{eq:W_update}, which steers power away from the eigendirections that couple most strongly into the ADCs, and from $\nu$ in~\eqref{eq:analog_LT}. Finally, projecting the analog taps onto $b_{\rm A}$-bit amplitude and phase vector modulators ($\alpha_{\rm r}=1$, $b=5$), with the digital canceller and the combiner re-optimized, costs at most $0.3$ and $0.4$~dB of radar SINR from $b_{\rm A}=5$ and $1.1$ and $1.3$~dB at $b_{\rm A}=4$, so the continuous analog canceller is a faithful model of a hardware one.

\begin{figure}[t]\centering\includegraphics[width=\columnwidth]{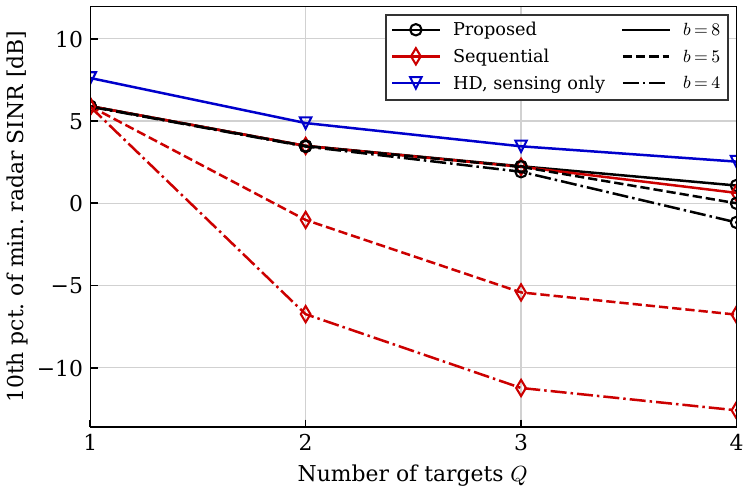}
\caption{10th percentile of the minimum radar SINR over $Q$ targets with equalized weights ($N=16$, $\alpha_{\rm r}=1$); HD reference $(\sum_q1/\bar\gamma_q)^{-1}$.}\label{fig:multi}\end{figure}

\subsection{Multiple Targets}

Fig.~\ref{fig:multi} considers $Q\in\{1,\ldots,4\}$ targets of equal RCS at random angles, $N=16$, $\alpha_{\rm r}=1$, and $b\in\{8,5,4\}$, with the target weights equalized by~\eqref{eq:fair_update} ($\kappa=0.3$, three inner iterations, $30$ outer steps) for both designs. The proposed design serves the targets on average within $0.2$~dB ($8$ bits) and $0.9$~dB ($4$ bits) of each other with a single multi-beam (the effective rank of $\bR_x$ is one in all but at most two of the $15$ realizations), for which $\xi_{\rm A}=0.1$ suffices (Corollary~\ref{cor:budget}). Its average minimum SINR with $8$-bit ADCs, $7.4$, $5.0$, $3.5$, and $2.4$~dB for $Q=1,\ldots,4$, follows the HD reference ($9.2$, $6.1$, $4.4$, and $3.2$~dB) at a distance set by the SI floor, and the sequential design performs similarly, since~\eqref{eq:P_adc} is inactive. With $5$- and $4$-bit ADCs, the constraint becomes active in $20$--$40\%$ of the realizations with $Q\ge2$, those in which the multi-beam aligns with a strongly coupled direction; the sequential design then backs off by up to $19.5$~dB and its weakest target falls to $-18.2$~dB, whereas that of the proposed design never falls below $-2.6$~dB. At $4$ bits and $Q=4$, the 10th-percentile minimum SINR is $-1.2$~dB against $-12.6$~dB, and the sequential design falls below $0$~dB in $33\%$ of the realizations, against $13\%$. Joint design is thus what makes multi-target sensing reliable with low-resolution ADCs.
\endgroup

\section{Conclusion}
\label{sec:conclusion}

We have developed a theory and a design for FD JCAS from a LEO satellite that exploit the separation, by tens of thousands of symbols, between the SI of the current transmission and the echo of a transmission several milliseconds earlier. The separation does not remove SI from the target range bin; what it changes is estimation. Restricting the cancellers to the short-delay SI support is without loss of optimality (Proposition~\ref{prop:support}), whereas a least-squares canceller whose support contains the target delay reduces the radar SINR by the factor $1-\zeta_q$, $17.8$~dB at the zero-Doppler crossing of a vessel (Lemma~\ref{lem:nd}, Corollary~\ref{cor:zeta}). \rev{When the non-reconstructible SI is spatially white,} the residual SI in the target bin floors the radar SINR at $\bar\gamma_q/(1+\epsilon^2\,\mathrm{INR})$ up to a negligible quantization factor (Proposition~\ref{prop:ceiling}\rev{; Section~\ref{sec:ceil} gives the bound for correlated SI}), a tap budget of about $r/N_t$ of the SI-channel energy suffices to reach this ceiling with a rank-$r$ transmit covariance (Lemma~\ref{lem:budget}, Corollary~\ref{cor:budget}), and the same floor sets the weight at which the optimal design changes over from communication to sensing (Proposition~\ref{prop:changeover}). The joint design is solved by an alternating algorithm with closed-form blocks, all canceller updates being instances of one generalized Sylvester solution. \revold{Simulations confirm the closed forms. With nominal hardware, a sequential design that keeps the SI-free beamformers and optimizes the cancellers for them reaches the trade-off curve of the joint design, but at the wrong weights, which costs up to $13\%$ in the normalized objective. Joint design matters where the hardware limits the cancellation: it gains $1$--$2$~dB of radar SINR with tight tap budgets and more than $10$~dB of downlink and radar SINR with $4$-bit ADCs and in the 10th-percentile worst-target SINR of four targets, whereas \rev{ignoring the SI overloads the ADCs}.}

\appendices

\section{Proof of Proposition~\ref{prop:support}}
\label{app:support}
Fix $\bW$ and $\{\bu_q\}$. The cancellers enter $\mathcal{J}$ through $\bR_{\rm A}^{\rm SI}$ in~\eqref{eq:RA}, through $\bR_{\rm SI}^{\rm res}$ in~\eqref{eq:Rres}, and through the range-compressed output~\eqref{eq:rc}. For $d\notin\Dsi$, $\bH_{\rm SI}[d]=\mathbf{0}$, so the residual channels at such delays are $\bHh_{\rm A}[d]=-\bC_{\rm A}[d]$ and $\bHh_{\rm res}[d]=-\bC_{\rm A}[d]-\bC_{\rm D}[d]$, and their contributions $\bC_{\rm A}[d]\bR_x\bC_{\rm A}^H[d]\succeq\mathbf{0}$ to $\bR_{\rm A}^{\rm SI}$ and $(\bC_{\rm A}[d]+\bC_{\rm D}[d])\bR_x(\cdot)^H\succeq\mathbf{0}$ to $\bR_{\rm SI}^{\rm res}$ vanish for the primed cancellers. If $d=d_q$ for some $q$, the subtracted term $-(\bC_{\rm A}[d_q]+\bC_{\rm D}[d_q])\bx[n-d_q]$ is correlated with the reference in~\eqref{eq:rc} and produces, after range compression, the deterministic term $-\sqrt{\Np/P_q}\,\bar\phi_q^*(\bC_{\rm A}[d_q]+\bC_{\rm D}[d_q])\bR_x\ba_t(\theta_q)$. Because $\E\{\rho_q\}=0$, this term is uncorrelated with the target component $\sqrt{\Np P_q}\rho_q\beta_q\ba_r$ and adds its power to the interference, not to the signal. Hence, for every $q$, the interference power $\bu_q^H\bR_{\rm i}\bu_q$ of the primed cancellers is no larger and the signal power is unchanged, so $\gamma_q$ does not decrease; $\gamma_k$ does not depend on the cancellers. Zeroing taps reduces $\sum_d\|\bC[d]\|_F^2$ and $\tr(\bR_{\rm A}^{\rm SI})$, so feasibility is preserved. \hfill$\blacksquare$

\section{Proof of Lemma~\ref{lem:nd}}
\label{app:nd}
Write the post-analog samples as $\by[n]=\rho_q\phi_q[n]\bH_q\bx[n-d_q]+\sum_{d\in\Dsi}\bHh_{\rm A}[d]\bx[n-d]+\mathbf{i}[n]$, where $\mathbf{i}[n]=\sum_{d\in\Dsi}\bDelta_n[d]\bx[n-d]+\bz[n]$ collects the non-reconstructible SI and the noise\revold{; the echo of another target is fitted like that of target $q$ if its delay lies in $\mathcal{S}$ and is otherwise uncorrelated with the regressors and with the reference, so it is omitted}; $\mathbf{i}[n]$ is zero mean and uncorrelated with $\bx[n-d']$ for every $d'$. By the law of large numbers, $\Np^{-1}\bX\bX^H\to\bI_{|\mathcal{S}|}\otimes\bR_x$ and $\Np^{-1}\bY\bX_d^H\to\E\{\by[n]\bx^H[n-d]\}$ time-averaged over the CPI. For $d\in\Dsi\cap\mathcal{S}$ this limit is $\bHh_{\rm A}[d]\bR_x$; for $d=d_q\in\mathcal{S}$, conditioning on $\rho_q$, it is $\rho_q\bar\phi_q\bH_q\bR_x$; for all other $d\in\mathcal{S}$ it is zero. Hence $\bC[d]$ converges to $\bHh_{\rm A}[d]\bR_x\bR_x^\dagger$, $\rho_q\bar\phi_q\bH_q\bR_x\bR_x^\dagger$, or $\mathbf{0}$, respectively, and $\bR_x\bR_x^\dagger$ is the projector onto the range of $\bR_x$, which contains $\bx[n]$ almost surely; this proves the convergence claims of (a) and (b).

For the range-compressed output, let $\mathbf{f}\in\Cbb^{\Np}$ with $[\mathbf{f}]_n=\phi_q^*[n]g_q^*[n-d_q]/\sqrt{\Np P_q}$, so that~\eqref{eq:rc} applied to a residual $\mathbf{E}_{\rm r}$ is $\mathbf{E}_{\rm r}\mathbf{f}$, and let $\bP_X=\bX^H(\bX\bX^H)^\dagger\bX$, so that the LS residual is $\bY(\bI-\bP_X)$. Since $g_q^*[n-d_q]=\bx^H[n-d_q]\ba_t$, the vector $[g_q^*[n-d_q]]_n$ lies in the row space of $\bX_{d_q}$. In case (b), write $\mathbf{f}=\bar\phi_q^*\mathbf{g}+\tilde{\mathbf{f}}$ with $[\mathbf{g}]_n=g_q^*[n-d_q]/\sqrt{\Np P_q}$ and $[\tilde{\mathbf{f}}]_n=(\phi_q^*[n]-\bar\phi_q^*)g_q^*[n-d_q]/\sqrt{\Np P_q}$; then $(\bI-\bP_X)\mathbf{g}=\mathbf{0}$, and $\bP_X\tilde{\mathbf{f}}$ has relative norm $\mathcal{O}(\sqrt{|\mathcal{S}|N_t/\Np})$ because $\Np^{-1}\sum_n(\phi_q^*[n]-\bar\phi_q^*)\bx[n-d]\bx^H[n-d_q]\to\mathbf{0}$ for every $d$. The reconstructible SI on $\mathcal{S}\cap\Dsi$ lies in the row space of $\bX$ and is removed. Hence the output is $(\mathbf{s}+\mathbf{I})\tilde{\mathbf{f}}$ up to the stated relative error, where $\mathbf{s}$ stacks the echo and $\mathbf{I}$ stacks $\mathbf{i}[n]$ together with the SI on $\Dsi\setminus\mathcal{S}$, if any, which is also uncorrelated with the reference. The echo gives $\sqrt{\Np P_q}\rho_q\beta_q\ba_r\,\Np^{-1}\sum_n\phi_q[n](\phi_q^*[n]-\bar\phi_q^*)=\sqrt{\Np P_q}\rho_q\beta_q\ba_r(1-\zeta_q)$, i.e., the power factor $(1-\zeta_q)^2$, whereas $\mathbf{I}\tilde{\mathbf{f}}$, being uncorrelated with the reference, has covariance $\Np^{-1}\sum_n|\phi_q[n]-\bar\phi_q|^2\,\bR_{\rm i}=(1-\zeta_q)\bR_{\rm i}$. In case (a), $\mathbf{f}$ is asymptotically orthogonal to the row space of $\bX$, so the output is $(\mathbf{s}+\mathbf{I})\mathbf{f}$ with the factors $1$ and $\bR_{\rm i}$. The ratio of the SINRs is $(1-\zeta_q)^2/(1-\zeta_q)=1-\zeta_q$, which proves~\eqref{eq:nd_loss}. \hfill$\blacksquare$

\section{Proof of Corollary~\ref{cor:zeta}}
For $\phi_q[n]=e^{j2\pi f_{\rm D}nT_s}$, $\bar\phi_q$ is a Dirichlet kernel, $|\bar\phi_q|=|\sin(\pi f_{\rm D}\Tcpi)|/(\Np|\sin(\pi f_{\rm D}T_s)|)\to|\sinc(f_{\rm D}\Tcpi)|$ as $\Np\to\infty$ with $\Tcpi$ fixed. For the chirp, $\bar\phi_q=\Tcpi^{-1}\int_{-\Tcpi/2}^{\Tcpi/2}\revold{e^{-j\pi f_{\rm R}t^2}}dt=\frac{2}{\Tcpi\sqrt{2f_{\rm R}}}\int_0^{X}\revold{e^{-j\pi u^2/2}}du$ with $u=\sqrt{2f_{\rm R}}\,t$ and $X=\sqrt{2f_{\rm R}}\,\Tcpi/2$; the integral is \revold{$C(X)-jS(X)$}, and $|\bar\phi_q|^2=4[C^2(X)+S^2(X)]/(2f_{\rm R}\Tcpi^2)=[C^2(X)+S^2(X)]/X^2$. The Doppler rate of a monostatic return at the zero-Doppler crossing is $f_{\rm R}=2v_{\rm eff}^2/(\lambda_cR_q)$, where $v_{\rm eff}$ is the geometric mean of the orbital and ground-track speeds. \hfill$\blacksquare$

\section{Proof of Proposition~\ref{prop:ceiling}}
\label{app:ceiling}
From~\eqref{eq:Ri}--\eqref{eq:Rres}, $\bu_q^H\bR_{\rm i}\bu_q\ge\sigma_r^2+\sigma_\Delta^2\tr(\bR_x)+\varrho\tr(\bR_{\rm A}^{\rm SI})$, and from~\eqref{eq:RA}, $\tr(\bR_{\rm A}^{\rm SI})\ge N_r\sigma_\Delta^2\tr(\bR_x)$. Hence $\bu_q^H\bR_{\rm i}\bu_q\ge\sigma_r^2+\sigma_\Delta^2P(1+\varrho N_r)$ with $P=\tr(\bR_x)$. For the numerator, $\bu_q^H\bR_q\bu_q=\sigma_q^2|\beta_q|^2|\bu_q^H\ba_r|^2\,\ba_t^H\bR_x\ba_t\le\sigma_q^2|\beta_q|^2N_r\,N_tP$ by the Cauchy--Schwarz inequality and $\ba_t^H\bR_x\ba_t\le\|\ba_t\|_2^2\lambda_{\max}(\bR_x)\le N_tP$. Therefore $\gamma_q\le\Np\sigma_q^2|\beta_q|^2N_rN_tP/[\sigma_r^2+\sigma_\Delta^2P(1+\varrho N_r)]$, which is increasing in $P$ and equals~\eqref{eq:ceiling} at $P=P_{\max}$. For the stated point, $\bHh_{\rm res}[d]\bR_x=(\bHh_{\rm SI}[d]-\bHh_{\rm SI}[d]\ba_t\ba_t^H/N_t)\ba_t\ba_t^HP_{\max}/N_t=\mathbf{0}$, so $\bR_{\rm SI}^{\rm res}=\sigma_\Delta^2P_{\max}\bI$, $\bR_{\rm A}^{\rm SI}=\sigma_\Delta^2P_{\max}\bI$, $\bR_{\rm i}\propto\bI$, and all inequalities hold with equality; the tap norm is $\sum_d\|\bHh_{\rm SI}[d]\ba_t\|_2^2/N_t$ and the pre-ADC power is $N_r\sigma_\Delta^2P_{\max}$, which gives the feasibility conditions. \hfill$\blacksquare$

\section{Proof of Proposition~\ref{prop:changeover}}
\label{app:changeover}
Write $\bw_k=c_k\bh_k+\mathbf{e}_k$ with $\bh_k^H\mathbf{e}_k=0$, and let $P_{\rm c}=\sum_k|c_k|^2\|\bh_k\|_2^2$ and $P=\tr(\bR_x)\le P_{\max}$. \revold{Dropping the interference in the denominator of $\gamma_k$, $\gamma_k\le|\bh_k^H\bw_k|^2/\sigma_k^2=|c_k|^2\|\bh_k\|_2^4/\sigma_k^2$}, so $\sum_k\gamma_k\le g_{\max}P_{\rm c}$ with $g_{\max}=\max_k\|\bh_k\|_2^2/\sigma_k^2$; \revold{maximum-ratio transmission of $P_{\max}$ to the strongest user attains this bound without SI, so} $\bar\Gamma_{\rm c}=g_{\max}P_{\max}$. For the target, $\ba_t^H\bh_k=0$ gives $\ba_t^H\bR_x\ba_t=\sum_k|\ba_t^H\mathbf{e}_k|^2+\ba_t^H\bR_{\rm s}\ba_t\le N_t(P-P_{\rm c})$, and, as in the proof of Proposition~\ref{prop:ceiling}, $\gamma_q\le\bar\gamma_q\,\frac{(P-P_{\rm c})/P_{\max}}{1+F P/P_{\max}}$ with $F=\epsilon^2\,\mathrm{INR}\,(1+\varrho N_r)$. Hence the normalized objective is at most \revold{$\alpha_{\rm c}\frac{P_{\rm c}}{P_{\max}}+\alpha_{\rm r}\frac{(P-P_{\rm c})/P_{\max}}{1+FP/P_{\max}}$}, which is linear in $P_{\rm c}\in[0,P]$ and therefore maximized at an endpoint, giving \revold{$\max\{\alpha_{\rm c}P/P_{\max},\,\alpha_{\rm r}(P/P_{\max})/(1+FP/P_{\max})\}$}; both terms increase with $P$, so the bound is \revold{$\max\{\alpha_{\rm c},\alpha_{\rm r}\bar\rho\}$; with $\alpha_{\rm c}=1-\alpha_{\rm r}$ the two terms are equal at $\alpha_{\rm r}^\star=1/(1+\bar\rho)$, and for $\alpha_{\rm r}\neq\alpha_{\rm r}^\star$} only the corresponding endpoint attains it. The HD-optimal beamformers give \revold{$\alpha_{\rm c}$}, and the point of Proposition~\ref{prop:ceiling} gives \revold{$\alpha_{\rm r}\bar\rho$}; both are feasible under the stated conditions, which proves attainment and~\eqref{eq:alpha_star}. \hfill$\blacksquare$

\section{Proof of Lemmas~\ref{lem:budget} and~\ref{lem:sylvester}}
\label{app:sylvester}
Problem~\eqref{eq:sylv_problem} is convex in $\{\bC[d]\}$ ($\bL\succeq\mathbf{0}$, $\bR_x\succeq\mathbf{0}$) with a convex constraint, and Slater's condition holds, so the KKT conditions are necessary and sufficient. Setting the gradient of the Lagrangian with respect to $\bC^*[d]$ to zero gives $\bL\bC[d]\bR_x+\vartheta\bC[d]=\bT_d\bR_x$. Substituting the eigendecompositions and $\widetilde{\bC}_d=\mathbf{E}^H\bC[d]\mathbf{F}$ decouples the equation entrywise, $([\bLambda]_{i,i}p_j+\vartheta)[\widetilde{\bC}_d]_{i,j}=[\widetilde{\bT}_d]_{i,j}p_j$, whose solution is~\eqref{eq:sylv_solution}; a zero denominator occurs only when $p_j=0$ or ($\vartheta=0$ and $[\bLambda]_{i,i}=0$), in which case the entry does not affect the objective and is set to zero for minimum norm. Since $\|\bC[d]\|_F=\|\widetilde{\bC}_d\|_F$ is strictly decreasing in $\vartheta$, complementary slackness fixes $\vartheta$ uniquely. Lemma~\ref{lem:budget} is the case $\bL=\bI_{N_r}$, $\bT_d=\bHh_{\rm SI}[d]$, for which the objective equals $\Pi_{\rm A}$ up to the constant $\sum_d\tr(\bHh_{\rm SI}[d]\bR_x\bHh_{\rm SI}^H[d])$; then $[\widetilde{\bC}_d]_{i,j}=[\widetilde{\bT}_d]_{i,j}p_j/(p_j+\vartheta)$, the residual in direction $j$ is $\sum_{d,i}|[\widetilde{\bT}_d]_{i,j}|^2p_j(1-p_j/(p_j+\vartheta))^2=h_jp_j\vartheta^2/(p_j+\vartheta)^2$, and the tap norm is $\sum_jh_jp_j^2/(p_j+\vartheta)^2$, which gives~\eqref{eq:pimin}. For Corollary~\ref{cor:budget}, full cancellation requires $\vartheta=0$, i.e., $\sum_{j:p_j>0}h_j\le P_{\rm A}^{\max}$; for eigenvectors generic with respect to $\bHh_{\rm SI}$, $\E\{h_j\}=\sum_d\|\bHh_{\rm SI}[d]\|_F^2/N_t$ by isotropy, which gives the approximation. \hfill$\blacksquare$

\balance
\bibliographystyle{IEEEtran}
\bibliography{refs}
\end{document}